%% file: main.tex
\documentclass[sigconf]{acmart}

\usepackage{makecell}
\usepackage{enumitem}
\usepackage{multirow}
\usepackage{longtable}
\usepackage{graphicx}
\usepackage{caption}
\usepackage{subcaption}
\usepackage{tabularx}
\usepackage{soul}
\usepackage{float}
\usepackage{url}
\definecolor{darkgreen}{rgb}{0.0, 0.5, 0.0}
\definecolor{navyblue}{rgb}{0.0, 0.0, 0.5}

\newif\ifredact
\newif\ifcomment

\redactfalse  
\commentfalse  

\newcommand{\dquote}[1]{\textit{``#1''}}
\newcommand{\ja}{\textit{joint attention}}
\newcommand{\pci}{\textit{Parent-Child Interaction}}
\newcommand{\ga}{\textit{gaze}}
\newcommand{\ac}{\textit{action}}
\newcommand{\vo}{\textit{vocalisation}}

\ifcomment
  \newcommand{\missing}[1]{\sethlcolor{yellow}\hl{[Missing: #1]}}
  \newcommand{\ken}[1]{\sethlcolor{yellow}\hl{[Kenny: #1]}}
  \newcommand{\wei}[1]{\sethlcolor{pink}\hl{[Weiyan: #1]}}
  \newcommand{\discuss}[1]{\sethlcolor{green}\hl{[Used in discussion: #1]}}
\else
  \newcommand{\missing}[1]{}
  \newcommand{\ken}[1]{}
  \newcommand{\wei}[1]{}
  \newcommand{\discuss}[1]{}
\fi

\AtBeginDocument{%
  }

\setcopyright{acmlicensed}
\copyrightyear{2018}
\acmYear{2018}
\acmDOI{XXXXXXX.XXXXXXX}

\acmConference[Conference acronym 'XX]{Make sure to enter the correct
  conference title from your rights confirmation emai}{June 03--05,
  2018}{Woodstock, NY}
\acmISBN{978-1-4503-XXXX-X/18/06}

\setcopyright{acmlicensed}

\copyrightyear{2026}
\acmYear{2026}
\setcopyright{cc}
\setcctype{by-nc-nd}
\acmConference[CHI '26]{Proceedings of the 2026 CHI Conference on Human Factors in Computing Systems}{April 13--17, 2026}{Barcelona, Spain}
\acmBooktitle{Proceedings of the 2026 CHI Conference on Human Factors in Computing Systems (CHI '26), April 13--17, 2026, Barcelona, Spain}
\acmPrice{}
\acmDOI{10.1145/3772318.3791267}
\acmISBN{979-8-4007-2278-3/2026/04}

\begin{document}

\title[Aligning Multimodal LLMs with Human Experts in Parent–Child Interaction]{Towards Aligning Multimodal LLMs with Human Experts:\\A Focus on Parent–Child Interaction}
\author{Weiyan Shi}
\email{weiyan\_shi@mymail.sutd.edu.sg}
\orcid{0009-0001-6035-9678}
\affiliation{
  \institution{Singapore University of Technology and Design}
  \city{Singapore}
  \country{Singapore}
}

\author{Kenny Tsu Wei Choo}
\email{kenny\_choo@sutd.edu.sg}
\orcid{0000-0003-3845-9143}
\affiliation{
  \institution{Singapore University of Technology and Design}
  \city{Singapore}
  \country{Singapore}
}
\renewcommand{\shortauthors}{Shi et al.}
\input{sections/0_abstract}


\begin{teaserfigure}
  \centering
  \includegraphics[width=0.6\textwidth]{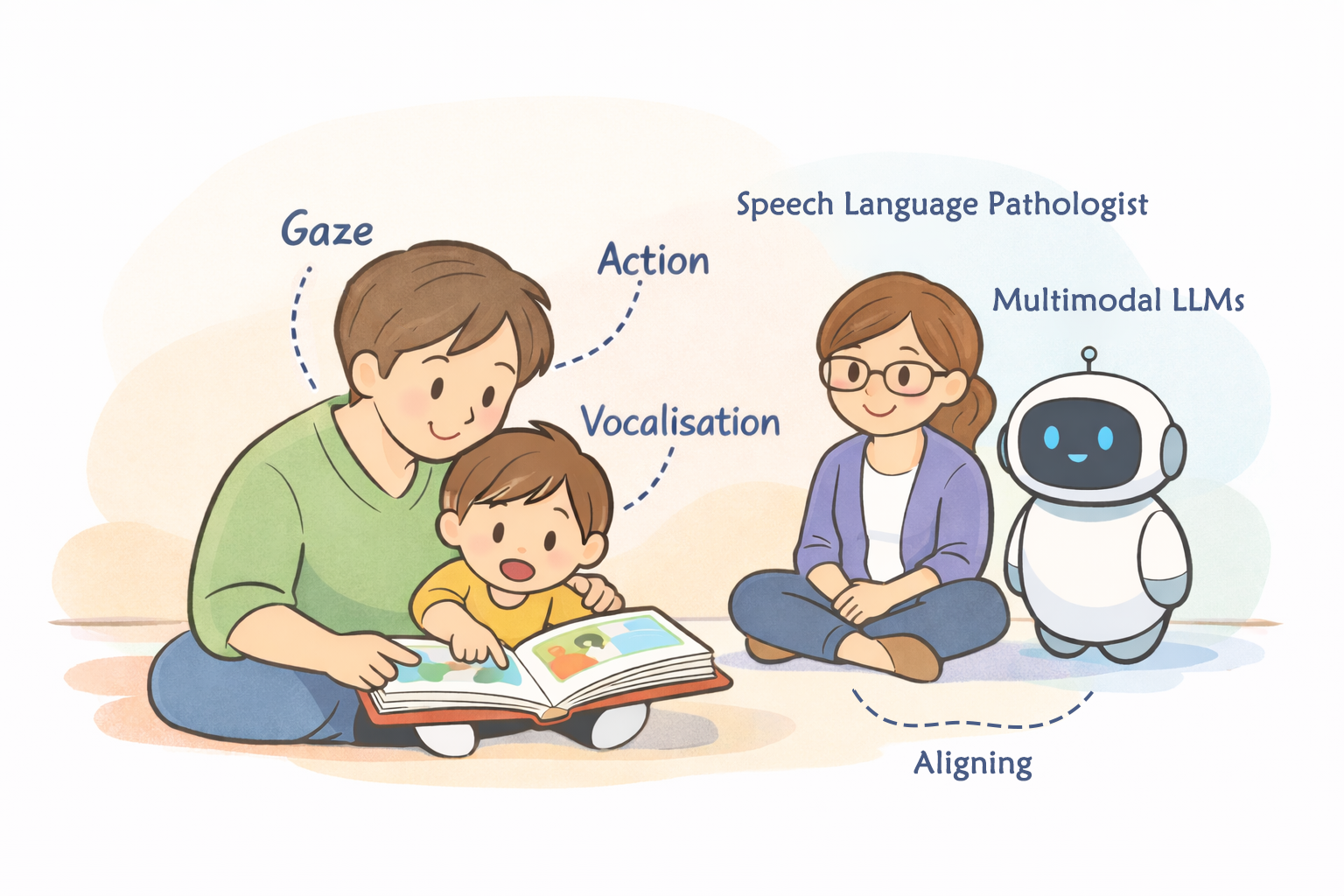}
  \caption{Conceptual overview: We investigate how multimodal LLMs can be guided to align with Speech-Language Pathologists in parent–child interaction through two key stages—Observation, using structured prompts based on behavioural cues such as gaze, action, and vocalisation; and Judgement, informed by few-shot learning with expert-derived practices.}
  \Description{This diagram illustrates how multimodal LLMs understands parent-child interactions by aligning with Speech-Language Pathologists and using cues like gaze, action, and vocalization.}
  \label{fig:banner}
\end{teaserfigure}


\maketitle


\input{sections/1_introduction}
\input{sections/2_relatedwork}
\input{sections/3_study}
\input{sections/4_mllmalignment}
\input{sections/7_discussion}
\input{sections/8_conclusion}


\bibliographystyle{ACM-Reference-Format}
\bibliography{main}

\input{sections/appendix}

\end{document}
\end{document}

%% file: sections/0_abstract.tex
\begin{abstract}
While multimodal large language models (MLLMs) are increasingly applied in human-centred AI systems, their ability to understand complex social interactions remains uncertain. We present an exploratory study on aligning MLLMs with speech–language pathologists (SLPs) in analysing joint attention in parent–child interactions, a key construct in early social–communicative development. Drawing on interviews and video annotations with three SLPs, we characterise how observational cues of gaze, action, and vocalisation inform their reasoning processes. We then test whether an MLLM can approximate this workflow through a two-stage prompting, separating observation from judgment. Our findings reveal that alignment is more robust at the observation layer, where experts share common descriptors, than at the judgement layer, where interpretive criteria diverge. We position this work as a case-based probe into expert–AI alignment in complex social behaviour, highlighting both the feasibility and the challenges of applying MLLMs to socially situated interaction analysis.
\end{abstract}

%% file: sections/1_introduction.tex
\section{Introduction}
Early parent-child interactions lay the foundation for lifelong communication, social, and cognitive development. 
Foundational social-communicative behaviours--such as joint attention, joint intention, and social referencing--emerge during play and serve as critical indicators of developmental progress. 
Yet these behaviours are often subtle and difficult for caregivers to identify without professional support. 
This challenge is consequential: approximately one in six children in the United States experiences at least one developmental delay~\cite{robinson2017cdc}, with language delays among the most prevalent~\cite{sunderajan_speech_2019}. 
At the same time, access to speech-language pathologists (SLPs)--professionals who assess children’s communicative development by attending to multimodal cues such as gaze, gesture, vocalisation, and context--remains limited, particularly in rural or underserved communities~\cite{o2005barriers}.
As a result, expert developmental knowledge and care often remain out of reach in everyday caregiving contexts.

Despite recent advances in artificial intelligence (AI) for family and education~\cite{yuan2024designing, sun2024exploring, fiani2024exploring, nikkhah2024family, currin2024opportunities, su2024hidden}, existing systems primarily assist with creating therapy materials or scaffolding caregiver engagement, rather than emulating the observing and judging practices of SLPs themselves.
For example, Lewis et al.~\cite{lewis2025exploring} investigated how SLPs interact with AI-generated materials for children from culturally and linguistically diverse backgrounds, and Dangol et al.\cite{dangol2025want} investigated how AI might support parents in delivering therapy at home, highlighting emotional and logistical challenges.
Yet in both cases, the evaluative expertise of SLPs remained external to the AI systems.
Little is known about whether MLLMs can be aligned to interpret parent-child interactions in ways that reflect expert reasoning.

In this paper, we explore how multimodal large language models (MLLMs) can be guided to emulate SLPs’ observing and judging practices.
We focus on joint attention as a representative case: a core developmental milestone that SLPs typically evaluate through holistic, experience-based interpretations of children’s gaze, actions, and vocalisations.
While MLLMs excel at structured tasks such as video captioning or temporal grounding~\cite{lu2024gpt}, they struggle with socially nuanced behaviours that lack clear ground truth.
At the same time, recent advances suggest that MLLMs can interpret complex verbal and non-verbal behaviour across video, audio, and text, and generate descriptive summaries of human activity~\cite{gandhi2023multimodal, jain2024vcoder}; however, their application to human behaviour analysis remains at an early stage.
MLLMs' ability to adapt to task-specific criteria through in-context learning makes them a promising candidate for bridging raw behavioural signals with the structured interpretive practices used by SLPs.
We therefore investigate how expert criteria can be translated into structured representations that allow MLLMs to more closely align with expert judgment.

Guided by this objective, we examine the following research questions (RQs):

\begin{itemize}
    \item \textbf{RQ1: How do SLPs observe and analyse parent-child interactions?} What criteria do they use to identify segments of strong or poor joint attention? How consistent are their judgments?
    \item \textbf{RQ2: How can we represent SLPs’ evaluative criteria in formats that MLLMs can both understand and execute, and what design choices best support alignment between MLLM and expert observation and judgement?}
\end{itemize}

To this end, we conducted in-depth interviews and annotation sessions with three experienced SLPs, who analysed 25 videos of parent-child interactions for strong and poor instances of joint attention.
From their annotations, we derived three key behavioural dimensions--\ga{}, \ac{}, and \vo{}--that underpin expert assessments.
We then translated these heuristics into structured, MLLM-compatible task formats: first prompting models to extract behavioural segments from raw video, and then prompting them to assess interaction quality using few-shot examples.

\textbf{Our key contributions are:}
\begin{itemize}
    \item We contribute an account of expert judgement practices around joint attention in parent-child interactions. Drawing on interviews and annotation analyses with SLPs, we show that their evaluations hinge on three interrelated cues: \textbf{\ga{}}, \textbf{\ac{}}, and \textbf{\vo{}}.

    \item We design and evaluate an exploratory MLLM system that aligns with speech-language pathologists’ approaches to joint attention assessment in two stages: (1) observing fine-grained behavioural cues from parent-child interaction videos using expert-informed prompting, achieving up to 85\% accuracy across dimensions; and (2) evaluating interaction quality using only structured behavioural descriptions, reaching over 64\% average accuracy compared to expert labels.

    \item We curate a segment-level dataset with expert-labelled joint attention behaviours and derive a set of practical design guidelines for building MLLM-based systems that align SLP’s observing and judging process. These guidelines cover prompt engineering, cue structuring, model configuration, and future directions for parent-facing AI systems.
\end{itemize}

%% file: sections/2_relatedwork.tex
\section{Related Work}

\subsection{Language Delay and the Limits of Current Parent Support}

Language delay is one of the most common developmental concerns in early childhood, with prevalence estimates ranging from 2.3\% to 19\% among children aged 2 to 7 years~\cite{mclaughlin2011speech}.
Without timely intervention, children with language delay are at increased risk for later difficulties in reading, attention regulation, and social interaction~\cite{sunderajan2019speech}.
SLPs play a key role in identifying and supporting children with language delays through structured interventions. 
However, limited access to professional services--whether due to long wait times, high cost, or service shortages--has led to growing interest in empowering parents to support their children at home~\cite{lieneman2017parent, straitstimes2024earlyintervention}.

Parent training programmes such as \textit{It Takes Two to Talk}, \textit{More Than Words}, and \textit{DIR Floortime} have shown success in encouraging behaviours that promote early communication and social development~\cite{pepper2004talk, sussman1999words, brock2014statewide, dir_floortime}.
These interventions typically guide caregivers to be more responsive and attuned during everyday interactions, and have demonstrated benefits in improving the quantity and quality of parent-child communication.
Yet while such programmes can teach parents strategies to encourage communication, they rarely equip caregivers to evaluate the developmental effectiveness of those interactions~\cite{finke2009all, gadberry2011survey, ganz2013impacts}.

For many parents, identifying nuanced developmental milestones can be difficult without professional guidance.
This is especially true for joint attention (\ja{}), one of the earliest and most important indicators of social-communicative development~\cite{mundy2007individual, tomasello1986joint}.
Determining whether \ja{} is strong, weak, or absent often requires the kind of interpretive expertise that SLPs develop over years of practice and is challenging to translate into formal rules~\cite{o2005barriers}.
Families outside of clinical environments rarely benefit from such expert judgment.
Motivated by this gap, we undertook this exploratory study oh how MLLMs might be aligned with SLPs' reasoning to assess joint attention in naturalistic parent-child interactions.
By probing the possibilities and limitations of this approach, we seek to inform future efforts to design AI tools that can augment professional expertise and broaden access to developmental support.

\subsection{Technological Support for Parent-Child Interaction}
There has been a surge in interactive technologies designed to support early childhood development, particularly by enhancing the quality and frequency of parent-child interaction. 
Early systems focused on lightweight support through manual logging.
For instance, Chan et al.~\cite{chan2017wakey} developed WAKEY, which aimed to smooth morning routines by prompting parents to record schedules and track phrase usage.
While effective in structuring communication, WAKEY relied entirely on manual input, without the ability to capture richer contextual signals.

Subsequent work shifted towards real-time speech analysis to provide parents with more immediate feedback. 
TalkBetter~\cite{hwang2014talkbetter} analysed turn-taking and generated tailored recommendations to scaffold language development, while TalkLIME~\cite{song2016talklime} extended this approach by offering live metrics such as utterance count, initiation ratio, and turn-taking frequency.

More recent systems incorporated multimodal sensing to deepen their understanding of interaction dynamics. Captivate!~\cite{kwon2022captivate} leveraged gaze estimation and speech recognition to detect joint attention and suggest parent prompts during play, and AACessTalk~\cite{choi2025aacess} supported minimally verbal autistic children by combining vocabulary recommendations with contextual guidance, ultimately improving turn-taking and parental reflection over a two-week deployment.

Parallel to these system developments, researchers have also explored how parents can be supported in adopting therapy strategies more effectively. Dangol et al.~\cite{dangol2025want} investigated how AI might address emotional and logistical barriers to at-home therapy, while Li et al.~\cite{li2025asd} introduced ASD-HI, a multimodal dataset and baseline model designed to detect strategies and assess fidelity in parent-led interventions.

Together, this trajectory reflects a shift from manual, parent-driven tools to increasingly automated, multimodal, and expert-aligned systems, yet few have investigated whether AI can evaluate interaction quality against professional standards--a gap our work explores by aligning MLLM-based analysis with SLP judgement.

\subsection{Technological Support for SLPs}
Generative AI is beginning to influence many aspects of clinical practice, including how SLPs source or adapt therapy materials.
For example, Lewis et al.~\cite{lewis2025exploring} examined how SLPs interact with AI-generated visual content when working with children from culturally and linguistically diverse backgrounds. 
While their study highlighted important concerns regarding representation, bias, and contextual mismatch, it focused primarily on content usability.
Their work does not address how AI might support or simulate the evaluative reasoning processes that SLPs use when observing and interpreting children’s behaviours in real time.
With the rapid progress of MLLMs across tasks such as video captioning and temporal grounding~\cite{lu2024gpt}, researchers have begun to explore their potential for analysing human behaviour--particularly in socially grounded contexts such as parent-child interaction.
MLLMs can interpret complex verbal and non-verbal behaviour across video, audio, and text, and generate descriptive summaries of human activity~\cite{gandhi2023multimodal, jain2024vcoder}. 
Thanks to their in-context learning capabilities, MLLMs can flexibly adapt to new interaction settings with minimal data~\cite{dos2023composite,whitehead2024generative}, making them promising for contexts that require domain-specific reasoning.
Crucially, these models shift the output from fixed metrics or visualisation into natural language descriptions--potentially bridging the gap between raw behavioural signals and expert interpretation.
Zheng et al.\cite{zheng2024soap} introduced SOAP.AI, a system that enables experts to interact with MLLMs through in-context prompting and collaborative task design, and supports the automatic generation of initial behavioural segments--such as those relevant to \ja{}--in domains like healthcare and education. However, SOAP.AI was designed primarily as a tool for expert users, without evaluating the reliability of its outputs or aligning them with SLP judgement. Extending this line of inquiry, Zheng et al.\cite{zheng2025ai} subsequently explored how AI-generated documentation could more directly support SLPs, identifying four opportunities and proposing three fluidity-focused design goals for future systems. However, while SOAP.AI exemplifies how AI can augment expert workflows, it is designed primarily for expert users and does not evaluate the reliability of its outputs or align them with SLP judgement, leaving open the question of whether MLLMs can simulate expert evaluative capabilities.

While these approaches centre on supporting SLPs in their clinical workflows, they do not address a critical question: can current technologies simulate the observational and interpretive judging that SLPs apply when evaluating parent-child interactions? 

While general-purpose systems such as Personalized Judge~\cite{dong2024can} show that LLM-as-Judge can be effective in broad settings, their limitations become clear in expertise-heavy domains. Szymanski et al.~\cite{szymanski2025limitations} and Cheng et al.~\cite{chen2024humans} show that although LLMs align moderately with subject-matter experts in broad fields such as dietetics and mental health (64–68\% agreement), their consistency drops on domain-specific questions that rely on professional judgement. Parent–child interaction illustrates this gap even more sharply: behaviours like joint attention require interpreting subtle, context-dependent cues rather than following fixed criteria. Prior attempts to prompt MLLMs for joint-attention grounding~\cite{shi2025towards} yielded near-chance performance, largely due to the models’ insensitivity to fine-grained gaze signals. These patterns suggest that, despite general promise, LLM-as-Judge remains insufficient for specialised behavioural assessment without deeper alignment to expert reasoning.

Our work continues this exploration by examining how MLLMs can be guided toward more SLP-aligned interpretations of parent-child interaction.

\begin{figure*}[ht]
    \centering
    \includegraphics[width=\linewidth]{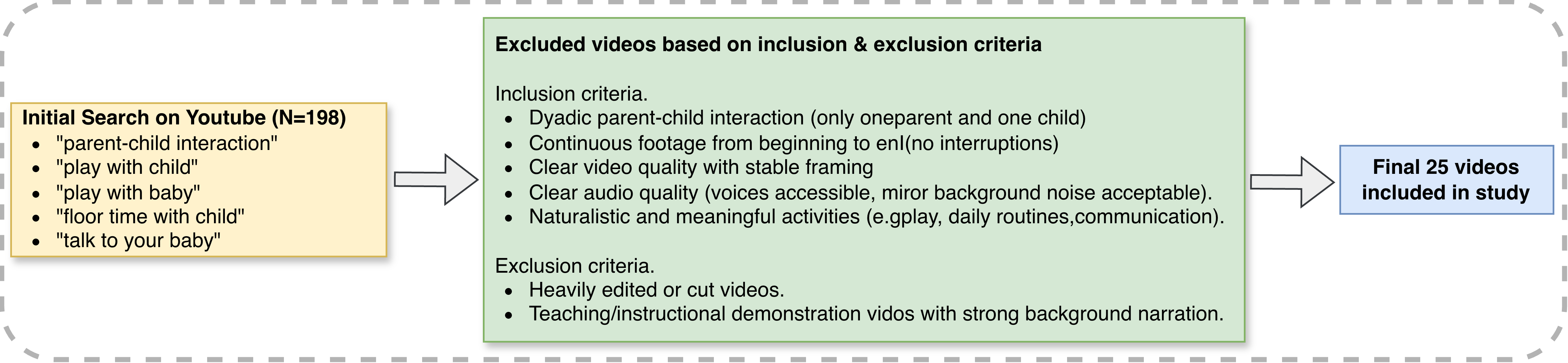}
    \caption{
    Flow chart of the video screening process
    }
    \Description{This diagram illustrates our video screening process, which began with a keyword search that yielded 198 videos. We then excluded videos based on our inclusion and exclusion criteria, resulting in a final set of 25 videos for the study.}
    \label{fig:screen}
\end{figure*}

%% file: sections/3_study.tex
\section{Exploratory Study of SLPs’ Understanding of Parent-Child Interaction Based on Joint Attention}
\label{sec:part1}

We began with a formative study to ground our exploration of how MLLMs might align with SLPs' practices.
First, we curated a corpus of 25 publicly available parent-child interaction videos from YouTube, which provided a diverse set of scenarios beyond the constraints of lab-based data collection.
We then conducted semi-structured interviews with three experienced SLPs to probe how they conceptualise \ja{} within \pci{} and to surface the behavioural cues they attend to when judging interaction quality.
Building on these discussions, the SLPs annotated selected video segments that exemplified both \textit{strong} and \textit{poor} instances of \ja{}, producing rich, expert-grounded labels that served as training and evaluation data for our MLLM-based system.
This study was reviewed and approved by the Institutional Review Board of Singapore University of Technology and Design (IRB approval no. IRB-24-00693), and all participants provided informed consent.
SLPs received approximately USD 31.2 as compensation for participation.

\subsection{Dataset}
We collected videos from YouTube using targeted search terms (e.g., \dquote{parent-child interaction}, \dquote{play with child}, \dquote{play with baby}, \dquote{floor time with child}, and \dquote{talk to your baby}) to capture demonstration-based home-setting parent-child interaction sessions.
Our goal was to find clear dyadic parent-child exchanges with continuous footage that flowed smoothly from beginning to end, allowing observation of how \ja{} behaviours emerge and develop.
Videos were required to provide clear behaviours, with stable framing and adequate audio, though minor background noise was acceptable. 
We excluded short or heavily edited clips and prioritised diverse, meaningful activities such as language learning, skill-building, and daily routines across a broad range of child ages. 
Elements such as subtitles or transitional animations were removed to preserve the raw parent-child interaction content. Figure~\ref{fig:screen} shows our video screening process. For the purposes of our study--aligning MLLM with human expertise, we noted that many of the selected videos were created by SLPs or professional organisations. Accordingly, in the subsequent stage, we recruited SLPs as our human experts.

We curated a final dataset of 25 parent-child interaction videos (Video 1-25) spanning developmental stages from infancy to early school age (0-8 years), with most focusing on preschool years.
Age information was obtained from video descriptions, covering children aged 0.5-2 (n=3), 2-4 (n=5), 4-6 (n=16), and 6-8 (n=1).
Videos ranged from 30 seconds to 12 minutes, though most were short (13 under 1 minute, 11 between 1-5 minutes). A full list of video IDs, titles, and source URLs is provided in Appendix~\ref{app:video-list}.
These videos were created and shared by parent-child interaction experts who have published numerous resources aimed at modelling effective interaction techniques and communication strategies for caregivers.
Given the difficulty of accessing authentic recordings online and the exploratory nature of this study, most of our dataset consists of practitioner-shared demonstration sessions, often created by certified SLP or PCIT trainers. 
These videos were openly accessible and chosen with the intention of enhancing accessibility at this early stage.

\begin{figure}[ht]
    \centering
    \begin{subfigure}[t]{0.31\textwidth}
        \includegraphics[width=\linewidth]{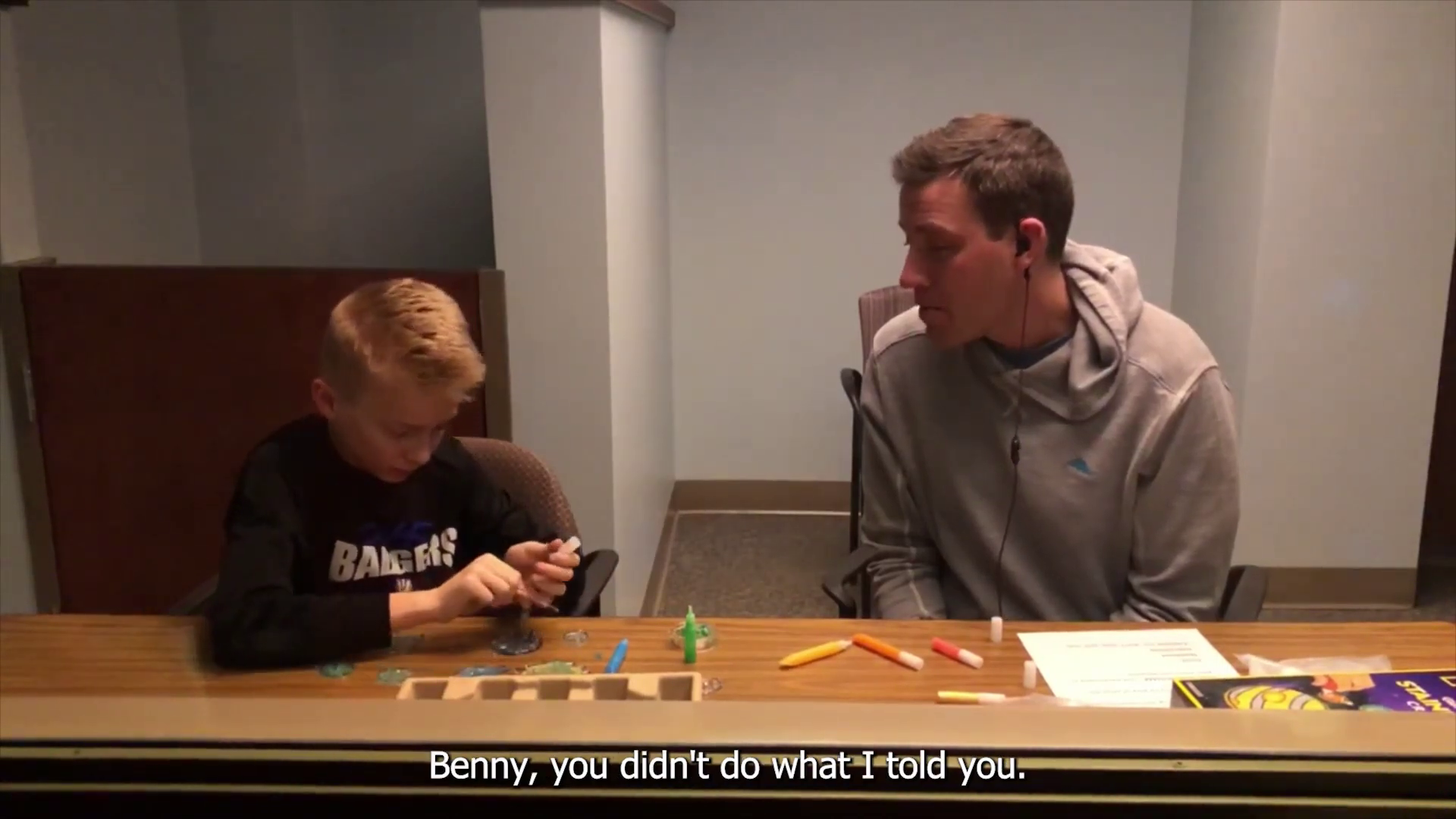}
        \caption{\textbf{Behavioural Guidance and Skill Modelling}: 
        Parent uses the \textit{"Big Ignore"}~\cite{woodfield2021time} technique while the child continues painting (Video~3~\cite{lieneman2023}).}
    \end{subfigure}
    \hfill
    \begin{subfigure}[t]{0.31\textwidth}
        \includegraphics[width=\linewidth]{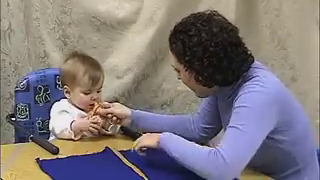}
        \caption{\textbf{Language and Cognitive Development}: 
        Parent and child play a toy-hiding game to foster engagement. (Video1~10~\cite{Lieneman2024})}
    \end{subfigure}
    \hfill
    \begin{subfigure}[t]{0.31\textwidth}
        \includegraphics[width=\linewidth]{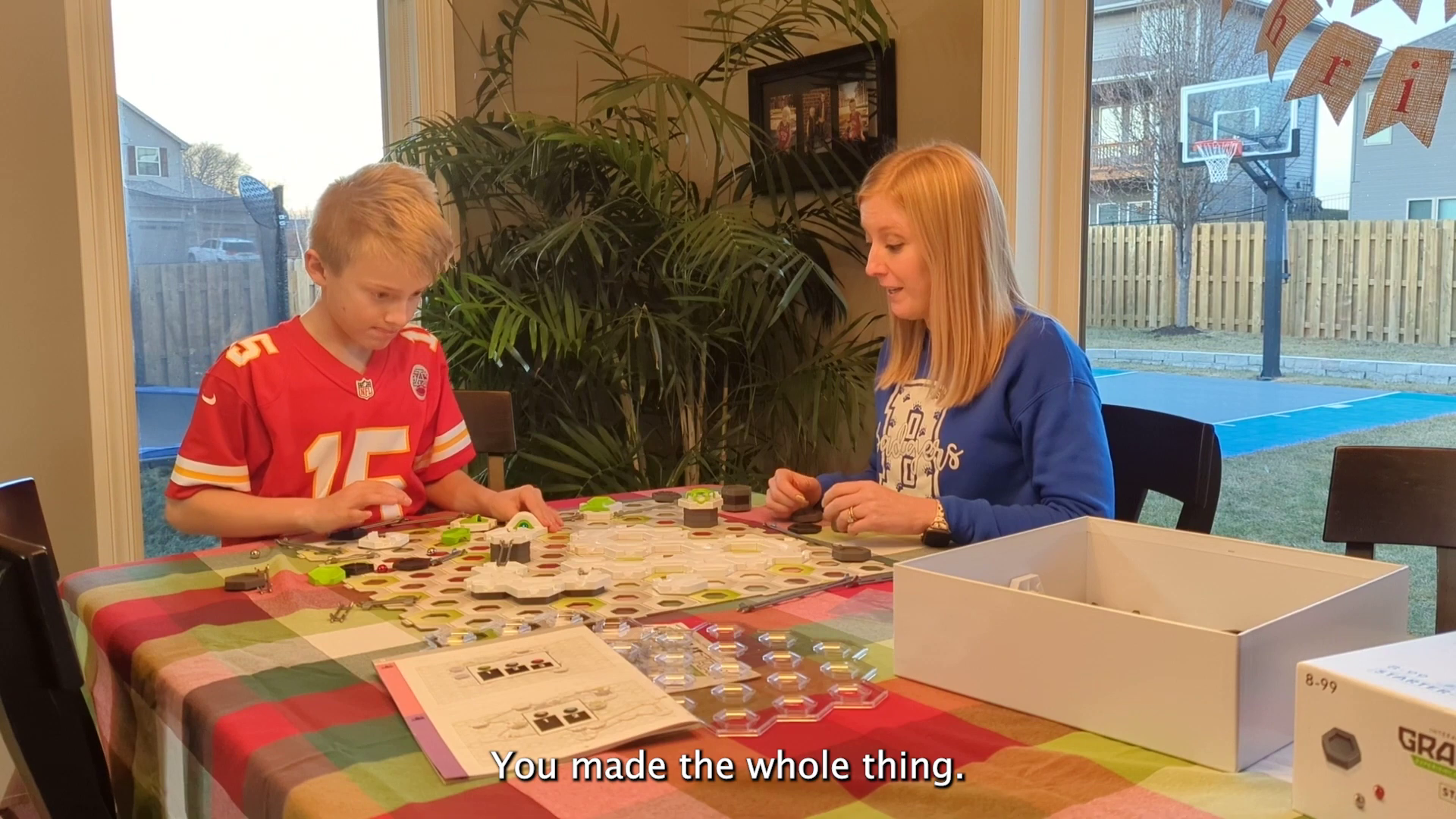}
        \caption{\textbf{Daily Life Skills and Interaction}: 
        Parent and child build a marble run together, encouraging planning (Video2~\cite{adam2013}).}
    \end{subfigure}
    \caption{Examples from three categories in our dataset: behavioural guidance, language development, and daily life interaction.}
    \label{fig:video_examples}
    \Description{This diagram shows example videos from our dataset, categorized into three types: Behavioural Guidance and Skill Modelling (left), where a parent uses the 'Big Ignore' technique while a child paints; Language and Cognitive Development (middle), where a parent and child play a toy-hiding game; and Daily Life Skills and Interaction (right), where they build a marble run together to encourage plannings}
\end{figure}

The videos in the dataset covers three categories: (1) behavioural guidance and skill modelling\footnote{\url{https://www.youtube.com/watch?v=YUkujhg6j6w}} (n=10), where parents demonstrate techniques such as \textit{PRIDE skills}~\cite{masse2018taking} or \textit{\dquote{Big Ignore}}~\cite{woodfield2021time}; (2) language and cognitive development\footnote{\url{https://www.youtube.com/watch?v=rVqJacvywAQ}} (n=9), showing tasks like topic discussions or Piaget’s experiments; and (3) daily life skills and interaction\footnote{\url{https://www.youtube.com/watch?v=N3wAPLXd7I0}} (n=6), including reward chart reviews or shared playtime.
Examples from each category are illustrated in Figure~\ref{fig:video_examples}.

\subsection{Participants} 

Our three expert SLPs were all female, aged between 30 and 59 years (see Table~\ref{tab:slpprofiles}). All have extensive clinical experience working directly with children, particularly those on the autism spectrum, and have received training in a wide range of evidence-based intervention programmes, including Hanen’s \textit{It Takes Two to Talk}~\cite{pepper2004talk}, \textit{More Than Words} and \textit{DIR Floortime}~\cite{dir_floortime}.

Their clinical practice spans public and private sectors and includes diverse settings such as early intervention centres, preschools, home-based care, and multidisciplinary teams. Notably, the SLPs have worked in different countries across multiple continents, contributing to a culturally informed understanding of parent-child interaction.

Given their specialised training and long-standing experience, particularly with children on the autism spectrum, who often experience joint attention difficulties--the SLPs were well-qualified to identify and evaluate strong and poor instances of joint attention.

\begin{table*}[ht]
    \centering
    \small
    \caption{Profiles of the three expert SLPs who participated in this study, including their age group, academic qualifications, specialist certifications, and experience years.}
    \begin{tabularx}{\textwidth}{llXX}
        \toprule
        \textbf{ID} & \textbf{Age Group}  & \textbf{Qualifications} & \textbf{Experience and Training} \\
        \midrule
        SLP1 & 30-34  & BSc in Speech Pathology; Certified in Hanen (\textit{It Takes Two to Talk}, \textit{More Than Words}); \textit{DIR Floortime} & 7 years of paediatric therapy experience in early childhood communication support, primarily in homes and SLP centres \\
        \midrule
        SLP2 & 30-34  & BSc in Speech Pathology; Certified in Hanen (\textit{It Takes Two to Talk}, \textit{More Than Words}); \textit{DIR Floortime}; \textit{PECS}; \textit{PROMPT}; \textit{Social Thinking}; \textit{TalkTools}; \textit{Key Word Sign Australia} & 9 years of paediatric therapy experience, including work with sensory processing and oral placement therapy approaches, primarily in SLP centres \\
        \midrule
        SLP3 & 55-59  & BSc in Speech Pathology; Msc in Communication Disorders; Certified in Hanen (\textit{It Takes Two to Talk}, \textit{More Than Words}); \textit{DIR Floortime} & 23 years of paediatric therapy experience, primarily in school settings, with extensive international practice experience \\
        \bottomrule
    \end{tabularx}
    \label{tab:slpprofiles}
\end{table*}

\subsection{Study Protocol}
\label{sec:stage1-protocal}

We conducted our study with each SLP individually.
The entire study was audio recorded.
The study comprises three stages and took approximately 2 hours:

\paragraph{\textbf{Pre-interview} (10 minutes)}
Each SLP was first asked to briefly share their professional background.
They were then asked to describe what \ja{} means in their professional judgment, what constituted strong, moderate, or poor \ja{}, the behavioural cues that they typically attend to, and how contextual factors may influence their evaluations.
They were also asked to share its significance for child development, and explain how \ja{} typically emerges and is supported through \pci{}.

\begin{figure}[h]
    \centering
    \includegraphics[width=\linewidth]{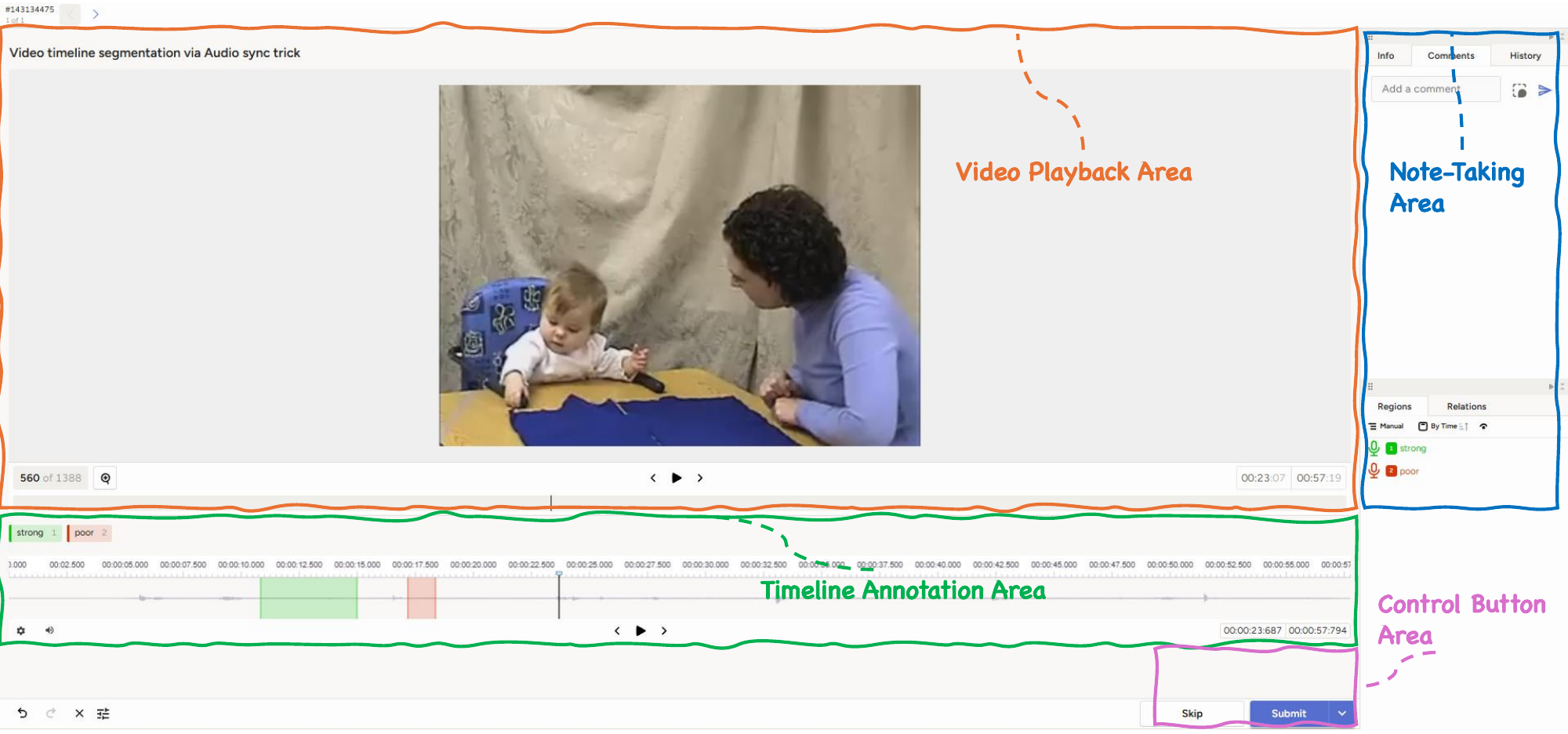}
    \caption{
    Our video annotation tool supports SLPs’ judgement process with four main components: 
    \textcolor{orange}{Video Playback Area} for watching, pausing, and replaying parent–child interactions 
    (shown here with a screenshot from Video~24.
    \textcolor{darkgreen}{Timeline Annotation Area} for selecting and labelling segments as \textit{strong} or \textit{poor} joint attention; 
    \textcolor{navyblue}{Note-Taking Area} for recording justifications or observations; and 
    \textcolor{violet}{Control Button Area} for task submission and navigation.
    }
    \label{fig:label-tool}
    \Description{This figure shows our video annotation tool, designed to support SLPs in their judgment process. It features four main components: a Video Playback Area (shown with a screenshot from Video 24) for watching interactions, a Timeline Annotation Area for labeling joint attention as "strong" or "poor," a Note-Taking Area for recording observations, and a Control Button Area for task submission and navigation.}
\end{figure}

\paragraph{\textbf{Annotation of \ja{}} (80 minutes)}
Each SLP then reviewed all 25 videos. 
To focus our analysis and ensure task feasibility, we limited annotations to the child's joint attention behaviour, which is often a primary concern in early developmental assessment.
They marked the start and end times of segments they considered to show \textit{strong} or \textit{poor} \ja{}, with all unlabelled segments treated as \textit{moderate} by default. 
Each annotation included their short explanation justifying their decision and their suggestions for how \pci{} could be supported.
We used our custom-built video annotation tool to facilitate the SLPs' judgement process (see Figure~\ref{fig:label-tool}, shown here with a screenshot from Video~24\footnote{\url{https://www.youtube.com/watch?v=rVqJacvywAQ}}~\cite{adam2013}).

\paragraph{\textbf{Post-task interview} (20 minutes)}
Finally, we revisited their annotation decisions to probe which observational cues (\ga{}, \ac{}, \vo{}) they relied on most, how they differentiated strong from poor instances.

\subsection{Data Analysis} 
The audio recordings were transcribed using WhisperX~\cite{bain2023whisperx} and manually verified.
We analysed the qualitative data by combining the SLPs’ transcripts from the interviews with their annotations during the video tasks.
To facilitate comparison across experts and to quantify areas of agreement and divergence, we standardised the data by dividing each video into fixed 5-second segments and mapping SLP labels onto these units. 
Any segments not explicitly labelled as \textit{Strong} or \textit{Poor} were automatically assigned a \textit{Moderate} label.

\subsection{Interview Study Results: SLPs’ Conceptual Understanding of Joint Attention}

In our semi-structured interviews, we sought to understand how each SLP conceptualises joint attention.
We asked them to define what joint attention means in their clinical judgment, how they differentiate between strong and poor instances, and whether they expect developmental differences in how joint attention is expressed across child age groups.
We summarised key insights from the interviews by grouping similar responses based on how the SLPs described and distinguished different forms of joint attention (see Table~\ref{tab:slp-concepts}).

\newcolumntype{L}[1]{>{\raggedright\arraybackslash}p{#1}}
\begin{table*}[ht]
\caption{SLPs' conceptual definitions of joint attention, their evaluative criteria, and their expectations during play across developmental stages.}
\centering
\small
\begin{tabularx}{\textwidth}{l L{3.5cm} L{3.5cm} L{4cm} L{4.5cm}}
\toprule
\textbf{ID} & \textbf{How They Define the Child's Joint Attention} & \textbf{What Counts as Strong vs. Poor} & \textbf{How Developmental Stage Affects Expression} & \textbf{How Joint Attention Appears During Play} \\
\midrule
SLP1 & 
Joint attention is the child’s ability to share focus and intent with another person, not just look at the same thing. It's about social connection. & 
Strong: back-and-forth referencing, shared interest \newline
Poor: no acknowledgment of other person’s presence & 
Yes -- younger children often rely more on gesture, posture, or affect due to limited language and attention control. For older children, SLP1 expects clearer behavioural signals and social reciprocity. & 
Joint attention fluctuates rapidly during play, especially in younger children. Distraction is expected and not inherently problematic. What matters is the overall balance--occasional poor moments are acceptable as long as strong episodes also occur. \\
\midrule
SLP2 & 
Joint attention is when a child follows or initiates shared gaze toward an object or action, and shows awareness that the other person is also attending. & 
Strong: gaze shifts, alternating eye contact \newline
Poor: no gaze response, ignores joint bids & 
Yes -- early on, gaze behaviour may be less consistent. By age 3, SLP2 expects reliable referential gaze, as cognitive and visual coordination typically improve with development. & 
Joint attention during play is highly variable and can shift within seconds. It’s normal for both strong and poor moments to co-occur within a short time window, such as within the same minute. \\
\midrule
SLP3 & 
Joint attention is a shared emotional and cognitive moment, often involving mutual engagement and affective alignment. It’s not just about behaviour but about intention. & 
Strong: emotional reciprocity, coordinated interaction \newline
Poor: child is disengaged despite cues & 
Yes -- SLP3 accounts for language, motor, and emotional maturity. For younger children, subtle affective signals may be enough. For preschoolers, SLP3 looks for intentional behaviours like pointing or verbal expression. & 
SLP3 believes it is unrealistic for a child to maintain strong joint attention throughout an entire play session. A child who always appears highly attentive--for example, staring continuously at the parent--would seem unusual. SLP3 described joint attention during play as something that \dquote{rises and falls like a temperature bar}, with natural shifts between strong and poor moments. What matters is the overall balance across the interaction, not perfection.\\
\bottomrule
\end{tabularx}
\label{tab:slp-concepts}
\end{table*}

These interviews revealed that while all three SLPs shared a common understanding of joint attention as a socially coordinated process, they differed in how they conceptualised and prioritised specific aspects of it. SLP1 emphasised the coordination of multiple communicative behaviours, and vocalisation--within dynamic, reciprocal interactions between child and parent. SLP2 focused more narrowly on whether the child performed clear gaze shifts between the adult and the shared object, treating visual referencing as the primary indicator of joint attention. In contrast, SLP3 viewed engagement as the core marker, paying particular attention to whether the child demonstrated emotional or attentional alignment with the adult, even in the absence of explicit cues like gaze or pointing.

All three experts agreed that joint attention can manifest differently across developmental stages, influenced by a child’s evolving language, motor, and attention capacities. For younger children--particularly those under age two--they allowed for broader interpretations, including reliance on gestures, posture, or affective responses. For older children, they expected more deliberate and referential signals, such as consistent gaze shifts, pointing, or verbal referencing, reflecting increased communicative intention and control.

Building on these conceptual insights, we next asked each SLP to annotate segments from our curated video dataset. This annotation study allowed us to observe how their stated definitions translated into actual evaluations of strong and poor joint attention in real-world parent-child interactions.

\subsection{Annotation Study Results: SLPs' Observational Judgements of Joint Attention}
During the annotation process, we observed that all three SLPs consistently relied on three core behavioural dimensions when evaluating joint attention. These dimensions--\textbf{\textit{gaze}}, \textbf{\textit{action}}, and \textbf{\textit{vocalisation}}--served as the primary basis for assessing both the child’s engagement and the parent's responsiveness. These cues were also frequently referenced in the SLPs’ verbal justifications, highlighting their prominence in expert reasoning. Below, we describe how each dimension was used in practice:

\begin{itemize}
    \item \textbf{\textit{Gaze}} -- Refers to where the child is looking, such as toward the parent's face, hands, a shared object, or away from the interaction. Gaze cues are critical for interpreting visual attention and shared focus.

    \item \textbf{\textit{Action}} -- Encompasses the child's physical behaviours, including pointing, reaching, standing up, dragging objects, or attempting to disengage. These actions signal communicative intent and social participation.

    \item \textbf{\textit{Vocalisation}} -- Includes all forms of vocal output, from babbling and laughter to verbal responses. Vocal cues provide insight into the child’s attempt to share attention or respond within the interaction.
\end{itemize}

This triadic lens emerged organically across SLPs and provided a shared framework for interpreting joint attention episodes in a structured yet flexible manner.

To facilitate structured analysis of expert annotations, we divided each video into uniform 5-second segments. While SLPs initially labelled joint attention quality by selecting start and end timestamps based on their judgement, we mapped these annotations onto the fixed segments, assigning each 5-second unit a corresponding label where overlap occurred. This mapping enabled us to translate variable-length judgements into a standardised format, making it easier to compare agreement across experts and analyse annotation patterns consistently across the dataset. Given that the activities--such as guided play or task-based exchanges--are generally short and focused, 5-second segments provided a practical and interpretable unit of analysis.

Table~\ref{tab:slp-distribution} summarises the distribution of joint attention labels assigned by each SLP, as well as the aggregated results based on majority agreement (i.e., at least two SLPs assigning the same label). Across all three experts, the majority of segments were labelled as \textit{Moderate}, suggesting that many observed behaviours fell into an intermediate range that did not clearly indicate strong or poor joint attention. Notably, \textit{Poor} labels were used relatively sparingly by SLP1 and only slightly more frequently by SLP2 and SLP3--reflecting a general reluctance to assign negative evaluations unless disengagement was highly evident.

The combined agreement row reveals that only 22.1\% of segments were jointly considered \textit{Strong}, and just 1.6\% were jointly labelled as \textit{Poor}, while over three-quarters (76.3\%) were consistently judged as \textit{Moderate}. This distribution highlights the interpretive nature of joint attention assessment and the need for careful alignment when designing systems to simulate expert reasoning.

To quantify the level of consistency among the three experts, we computed inter-annotator agreement using Krippendorff's $\alpha$ with an ordinal scale, which reflects the ordered nature of the \textit{Strong}, \textit{Moderate}, and \textit{Poor} labels. The overall $\alpha$ was 0.19, indicating low consistency among the three SLPs. In combination with the heavily skewed label distribution, where each of the three experts marked more than 60\% of segments as \textit{Moderate}, the result suggests that disagreements often arose in borderline cases rather than in clearly distinct behaviours.

Pairwise analyses further reveal asymmetries in expert alignment. The highest agreement occurred between SLP1 and SLP2 ($\alpha = 0.31$), while substantially lower agreement was observed for SLP1–SLP3 ($\alpha = 0.15$) and SLP2–SLP3 ($\alpha = 0.11$). These differences suggest that the three SLPs do not form a single coherent “consensus group”; instead, their judgments cluster into at least two partially aligned perspectives. Taken together, the low overall $\alpha$ and uneven pairwise consistency indicate that expert interpretation of joint attention is inherently variable, especially in segments where behaviours fall between categories. This variability reinforces the importance of modelling expert diversity rather than assuming a single ground truth.

\subsubsection{Illustrative Examples of Consensus Among SLPs}

While expert disagreement was common across many ambiguous segments, a smaller set of segments achieved full agreement among all three SLPs. These moments of consensus reveal the behavioural patterns that experts collectively interpret as clear evidence for strong, moderate, or poor joint attention. In general, jointly labelled \textit{Strong} segments featured multiple forms of engagement--such as mutual gaze, verbal responsiveness, and shared task focus--while \textit{Poor} segments were marked by disengagement, lack of social referencing, or solitary behaviour. Meanwhile, segments labelled \textit{Moderate} often involved task participation without clear signs of shared social coordination. 
The following examples help clarify the implicit thresholds that delineate high-confidence assessments in expert judgement:

\begin{itemize}
    \item \textbf{Video 19 Segment 001}: In this segment, the parent placed coins on the table while explaining the setup (“So I’m going to make two rows of quarters”) and directed their gaze to the coins. The child, seated with hands clasped on the table, briefly vocalised (“oh”) and followed the parent’s actions with their gaze. The combination of attentiveness, vocal response, and shared focus led all three SLPs to rate the segment as \textit{Strong}.
    \item \textbf{Video 2 Segment 004}: In this segment, the parent sat on the floor, picked up Lego blocks, and labelled them by saying “Tower.” The child, lying on the floor, manipulated the blocks and verbally speculated about their play (“Oh, it could be a…”). Both parent and child kept their gaze on the Lego rather than each other, resulting in limited reciprocity. All three SLPs rated the segment as \textit{Moderate}.
    \item \textbf{Video 9 Segment 001}: In this segment, the parent sat on the floor facing the child and offered praise (“Ooh, I like the way you’re connecting those guys”) while looking toward the child and the toys. The child, kneeling on the floor and connecting toy pieces, kept their gaze down on the toys and did not respond verbally. The lack of reciprocity or shared focus led all three SLPs to rate the segment as \textit{Poor}.
\end{itemize}

\begin{table}[ht]
\centering
\caption{Distribution of Joint Attention labels across individual SLPs and the combined agreement set ($\geq$2 SLPs). Each SLP annotated all 638 five-second video segments. The combined agreement set excludes segments without at least two SLPs agreeing on the same label, resulting in 615 segments (23 segments removed due to lack of consensus).}
\begin{tabular}{p{1.5cm}cp{1.5cm}p{1.5cm}p{1.5cm}}
\toprule
\textbf{ID} & \textbf{Total} & \textbf{Strong (n,\%)} & \textbf{Moderate (n,\%)} & \textbf{Poor (n,\%)} \\
\midrule
SLP1 & 638 & 155 (24.3\%) & 472 (74.0\%) & 11 (1.7\%) \\
SLP2 & 638 & 195 (30.6\%) & 392 (61.4\%) & 51 (8.0\%) \\
SLP3 & 638 & 150 (23.5\%) & 437 (68.5\%) & 51 (8.0\%) \\
\midrule
\textbf{Combined ($\geq$2 agree)} & \textbf{615} & \textbf{136 (22.1\%)} & \textbf{469 (76.3\%)} & \textbf{10 (1.6\%)} \\
\bottomrule
\end{tabular}
\label{tab:slp-distribution}
\end{table}

These consensus segments demonstrate how, despite individual stylistic differences, experts converge when multiple cues co-occur (for \textit{Strong}), are present but non-social (for \textit{Moderate}), or are entirely absent (for \textit{Poor}). Such segments serve as useful anchors when training or evaluating AI models that aim to simulate SLP-like assessments.

\subsubsection{Illustrative Examples of Diverging SLP Judgements}

Although all three SLPs shared a general understanding of joint attention as a coordinated social process, their specific criteria and interpretive emphasis varied. These differences became especially evident when they encountered ambiguous or intermediate segments.

\textbf{SLP1} adopted a balanced and multimodal approach. While gaze was still considered, it was treated as one of several contributing cues--alongside gestures, verbal referencing, emotional affect, and body orientation. SLP1 often acknowledged child-led play and nonverbal forms of engagement, showing flexibility in how joint attention could be expressed. Segments with inconsistent gaze but strong affective presence were still given high ratings when the child’s overall behaviour reflected social coordination.

\textbf{SLP2}, in contrast, placed strong emphasis on visual engagement, particularly gaze alternation between the adult and shared object. Eye contact and sustained visual referencing were seen as essential for identifying strong joint attention. In the absence of clear gaze cues, even segments involving verbal speech or physical interaction were often rated as moderate or poor. For example, SLP2 rated several segments as poor when the child appeared engaged but did not meet the baseline requirement of mutual gaze.

\textbf{SLP3} took a functional and context-sensitive perspective, often evaluating segments based on the child’s communicative intent rather than adherence to typical behavioural markers. Rather than expecting conventional cues like pointing or verbalisation, SLP3 recognised alternative forms of participation--such as pushing away the parent’s hand or physically repositioning themselves--as valid signs of engagement. This expert was less concerned with surface-level eye contact and more attentive to whether the child demonstrated awareness and responsiveness in their own way.

To illustrate these differences, we highlight several segments where the three SLPs disagreed in their assessments. These examples underscore the unique interpretive lenses each expert applied:

\begin{itemize}
    \item \textbf{Video 11 Segment 002}: In this segment, the child sat on the floor playing with a toy airplane and said, \dquote{I want to go to the Bahamas. Fly there now.} While there was no gaze toward the parent, the child’s speech was contextually rich. SLP1, recognising the imaginative verbal output and engagement, rated it as \textit{Moderate}. SLP2 labelled the segment as \textit{Poor} due to the lack of visual coordination. SLP3 rated it \textit{Strong}, interpreting the child’s verbal and motor actions as clear evidence of communicative intent.
    \item \textbf{Video 21 Segment 040}: The child alternated gaze between a toy and the parent’s hands while reaching for the toy, but did not speak. SLP1 gave a \textit{Moderate} rating, due to the absence of vocal engagement. SLP2 labelled this as \textit{Strong}, citing the clear visual coordination. SLP3 judged it as \textit{Poor}, arguing that the interaction lacked reciprocal cues or intentional signalling.
    \item \textbf{Video 21 Segment 017}: In this segment, the child pointed to a toy lion without speaking. SLP1 rated it \textit{Moderate}, perhaps noting the absence of emotional or verbal expression. SLP2 rated the interaction as \textit{Strong}, identifying the combination of gaze and pointing as sufficient for joint attention. SLP3 gave it a \textit{Poor}, interpreting the gesture as mechanical rather than socially directed.
\end{itemize}

Together, these examples demonstrate how gaze, vocalisation, and action were weighted differently by each expert, and how their background and theoretical orientation shaped their interpretation. This variability in interpretive emphasis directly aligns with the earlier statistical results, where overall inter-annotator agreement remained low ($\alpha = 0.19$), reflecting the inherent subjectivity of these judgements.

%% file: sections/4_mllmalignment.tex
\section{Exploring Aligning MLLMs with SLPs: Behaviour Observations to Judgement}
\label{sec:part2}

Building on our findings from the interview and annotation studies, we sought to model SLPs' reasoning processes computationally.
We designed a two-stage system that enables MLLMs to simulate how SLPs observe and judge joint attention.
The first stage focuses on \emph{structured behavioural observation}--using the three expert-derived dimensions of \textit{gaze}, \textit{action}, and \textit{vocalisation} to prompt the MLLM to describe what is happening in each video segment. 
The second stage then assesses whether these structured descriptions improve the MLLM’s ability to judge the quality of joint attention. We evaluate this by comparing zero-shot and many-shot prompting strategies, and contrasting reasoning-based and non-reasoning model variants. Through this pipeline, we examine how well MLLMs can align with expert criteria in both perception and judgement.

\subsection{Stage~1: Behaviour Description through Expert-Aligned Prompting}

\subsubsection{Method}

With the rapid advancement of MLLMs, their application to human behaviour analysis presents a promising frontier.
In this study, we adopt \texttt{Gemini 2.5 Pro}\footnote{\url{https://blog.google/products/gemini/gemini-2-5-pro-updates/}} for its strong multimodal reasoning capabilities~\cite{yue2024mmmu, team2024gemini, fu2024video}.
A further practical consideration is that Gemini 2.5 Pro is, at present, the only widely accessible API model that accepts direct video inputs. Other foundation models, including the latest OpenAI offerings, currently support only image-based visual inputs, which makes them less suitable for segment-level behavioural captioning in continuous parent-child interaction videos. However, prior research~\cite{shi2025towards} shows that directly prompting MLLMs to identify key interaction segments--especially for complex, socially embedded tasks like joint attention--often fails. Models continue to struggle with long video understanding, temporal grounding, and fine-grained cues such as gaze interpretation.

To address these limitations, we build on insights from our SLP interviews, which revealed two important features of joint attention evaluation: (1) key behavioural changes are rapid and often context-independent, and (2) expert assessments consistently rely on three observable cues--\textbf{gaze}, \textbf{action}, and \textbf{vocalisation} (see Table~\ref{tab:slp-concepts}).

Guided by these principles, we divided each video into uniform 5-second segments and designed a structured prompting scheme to elicit behavioural summaries from the MLLM. For each segment, the model was asked to describe both the parent and the child’s behaviour along the three core dimensions. To encourage accurate, grounded behavioural descriptions, we designed a zero-shot prompt that guides the model to produce short, factual observations in natural language. By enforcing a consistent subject-verb-object structure, the prompt reflects the interpretive style of human annotators and reduces the likelihood of hallucinated or overly abstract outputs. The full prompt is shown below:

\noindent\textbf{\textit{Structured Behaviour Description Prompt\\(Gaze-Action-Vocalisation)}}:
\begin{quote}\ttfamily
\small
You are watching a video of a parent interacting with a child.

For each participant (parent and child), describe their behaviour in three parts:

1. Gaze: Describe what or whom the person was looking at, using a natural language sentence with a subject-verb-object structure.  
   Examples:  
   - The child looked at the parent’s face.  
   - The parent shifted gaze between the child and the toy.  
   - The child stared at the blocks.  
   - The parent looked away for a moment.

2. Action: Describe what the person physically did, using a short natural language sentence with a subject-verb-object structure.  
   Examples:  
   - The parent pointed at the red ball.  
   - The child reached towards the puzzle pieces.  
   - The parent lifted the toy truck.  
   - The child clapped their hands.

3. Vocalisation: Transcribe or paraphrase what the person said, or describe any vocalisations using a short sentence with a subject-verb-object structure.  
   Write `None' if there were no vocalisations.
\end{quote}

\subsubsection{Result}
We ran the full set of behaviour description prompts on an Ubuntu 22.04 LTS server via the official \texttt{Gemini 2.5 Pro} API, collecting MLLM-generated outputs for all 5-second video segments in our dataset.

To evaluate the MLLM’s behavioural description performance, we first conducted a structured human review of all model-generated outputs for each segment. Two annotators from our research team independently examined every 5-second segment. For each segment, they reviewed all six model-generated entries (parent/child $\times$ gaze/action/vocalisation) and assigned a binary correctness label: \textbf{1} if the description accurately matched the observed behaviour in the video, and \textbf{0} otherwise. For each of the three behavioural fields, correctness was determined according to the subject--verb--object structure specified in the prompt. For \textit{gaze}, an entry was marked as correct only if both the subject and the gaze target (object) were accurate, such as distinguishing between looking at the parent's face or looking at the blocks held in the parent's hand. For \textit{action}, correctness required that the subject and the action described were consistent with the video, for example identifying whether the child was using a hand to pick up a block or performing another distinct motor behaviour. For \textit{vocalisation}, an entry was considered correct only when the described verbal content matched what was actually spoken in the segment, including both who produced the speech and the approximate content of the utterance. These field-specific criteria ensured that the binary labels reflected the accuracy of the subject--verb--object components rather than surface-level wording.

The two annotators achieved an inter-annotator agreement of Cohen's~$\kappa = 0.92$, indicating high consistency. Any disagreements were resolved through discussion, and most arose from fine-grained distinctions in gaze descriptions. For example, the model sometimes output \dquote{child look at parent} when the behaviour should instead specify whether the child was looking at the parent's face or hand. Based on these human-verified binary correctness labels, we computed accuracy for each behavioural field at the level of entire videos. For each video, we aggregated all segment-level binary labels for that field and calculated a per-video accuracy score. Table~\ref{fig:mllm} reports the mean, median, maximum, and minimum of these per-video accuracies across the full dataset. As shown in the table, the MLLM performed best in the \textit{action} category, with a mean per-video accuracy of 0.88, followed by \textit{vocalisation} (0.87) and \textit{gaze} (0.86). All three fields achieved perfect accuracy in at least one video, although \textit{gaze} also had the lowest minimum accuracy, highlighting its difficulty for the model.

\begin{table}[h]
\centering
\caption{Per-video aggregated accuracy statistics. For each field, accuracies were computed per video, 
and the table summarises the mean, median, maximum, and minimum across all videos.}
\begin{tabular}{lcccc}
\toprule
\textbf{Field} & \textbf{Mean} & \textbf{Median} & \textbf{Max} & \textbf{Min} \\
\midrule
Action        & 0.8774        & 0.9464          & 1.0000       & 0.6250       \\
Vocalisation  & 0.8708        & 0.9259          & 1.0000       & 0.5000       \\
Gaze          & 0.8556        & 0.8750          & 1.0000       & 0.5000       \\
\bottomrule
\end{tabular}
\label{fig:mllm}
\end{table}

To better understand model failure cases, we conducted a qualitative review of low-performing videos across the three behavioural fields. Several recurring issues emerged:
\begin{itemize}
  \item \textit{Speech role confusion} was a major factor affecting vocalisation accuracy. For instance, in \textit{Video~24} and \textit{Video~25}, the model consistently attributed child-like vocalisations to the parent-particularly when the parent mimicked the child’s babbling. Similar confusion was observed in \textit{Video~11}, where rapid turn-taking and overlapping speech made speaker attribution unreliable.
  
  \item \textit{Gaze misinterpretation} often occurred when faces were partially occluded, or when the child looked at non-face targets such as hands or objects. In \textit{Video~11}, the child’s gaze was repeatedly marked as \dquote{looking at the parent} despite clear visual evidence that the child was focused on the toy airplane.

  \item \textit{Action detection errors} were more frequent in unstructured scenes. For example, in \textit{Video~4}, the child’s aggressive toy-hitting was not recognised, and in \textit{Video~15}, a clear tray-passing motion was entirely missed. In \textit{Video~5}, task misunderstanding led to mismatched labels (e.g., writing mistaken for eating).
\end{itemize}

These results demonstrate that our structured prompting approach substantially improved the reliability of MLLM-generated behavioural annotations, enabling the model to produce interpretable outputs that aligned with expert-labelled segment judgments in many cases. Shi et al.~\cite{shi2025towards} first introduced this behavioural description task and dataset, reporting that the model achieved approximately 62\% behavioural accuracy within the 23\% of segments where it successfully detected gaze cues. Building on this prior setup, our structured prompting approach ensures that the model receives explicit action, vocalisation, and gaze information for every segment and thus produces complete descriptions for 100\% of the dataset. Using a comparable combined accuracy metric that marks a segment as correct only when all three dimensions are correct, our method achieved 81\% accuracy. This reflects a substantial improvement in both overall description quality and gaze-related interpretation.

\subsection{Stage~2: Challenges in Simulating SLP Joint Attention Judgement with Prompting}

In Stage 2, we investigated whether MLLMs could approximate SLPs' judgements of joint attention when provided with the behavioural cues identified in Stage 1.
This step shifts from describing observable behaviours to making evaluative, subjective judgements--a task that is inherently more challenging.
We conducted an experiment with \texttt{GPT-4.1}, comparing zero-shot and many-shot prompting strategies, and evaluated model outputs against SLP annotations using \textit{precision}, \textit{recall}, and \textit{F1}.

\subsubsection{Dataset}

The Stage~2 dataset was derived from the model-generated segment descriptions produced in Stage~1. For each 5-second segment across the 25 parent--child interaction videos, the model output six behavioural entries (parent/child $\times$ gaze/action/\\vocalisation). Two annotators independently evaluated the correctness of each entry based on the video, and any disagreements were resolved through discussion until consensus was reached. This consensus served as the ground truth for determining which descriptions required correction.

We then applied targeted revisions only to the entries marked as incorrect during the agreement process. Following a minimal-editing principle, we retained the original subject--verb--object structure whenever possible and corrected only the behavioural element that was factually inaccurate or mis-specified (e.g., replacing \textit{“the child looked at the parent”} with \textit{“the child looked down at the table”} if no eye contact occurred). When the model’s description was underspecified but directionally correct, we refined it to align with SLP-like granularity, such as specifying the actual attentional target (e.g., \textit{“the child looked at the parent’s hand”}).

The resulting set of minimally corrected and consensus-verified descriptions constituted the input to the Stage~2 judging process.\footnote{See the full dataset at \url{https://github.com/weiyan-shi/PCI-Align}.}

\subsubsection{Experiments}

For the Stage~2 judging process, we used \texttt{GPT-4.1 (gpt-4.1-2025-04-14)}\footnote{https://platform.openai.com/docs/models/gpt-4.1}
 as the backbone model for all text-based decisions. Stage~2 mirrors how SLPs conduct holistic assessment: instead of observing isolated 5-second segments, the model must read the entire sequence of segment-level descriptions from a full video and determine which portions correspond to \textit{Strong}, \textit{Moderate}, or \textit{Poor} joint attention. 
 This requires integrating behavioural evidence distributed across the interaction, similar to how an SLP reviews the full temporal flow of gaze, action, and vocalisation before forming a judgement. The task therefore primarily requires stable long-context understanding, enabling the model to integrate behavioural descriptions that span the full interaction rather than treating segments independently.

Although no existing LLM-as-judge benchmark directly evaluates this form of segment-level behavioural scoring, the Stage~2 task imposes requirements that differ substantially from Stage~1. Stage~1 relies on Gemini~2.5~Pro, a strong video–reasoning model that excels at extracting behavioural cues from short, self-contained 5-second clips. In contrast, Stage~2 mirrors how SLPs make holistic judgements: the model must read the entire sequence of segment descriptions for a full video, integrate behavioural evidence distributed across many segments, and determine which portions correspond to \textit{Strong}, \textit{Moderate}, or \textit{Poor} joint attention. In our dataset, the longest 12-minute video produced approximately 60{,}000 tokens of segment-level descriptions, meaning the Stage~2 model must process and integrate an input roughly of this size for a single judgement. As a result, long-context reliability, rather than step-by-step reasoning, becomes the primary requirement for Stage~2.

At the time of our experiments, GPT-4.1 was the strongest publicly available model for long-context understanding. It supports a verified 1M-token context window and demonstrates superior performance compared with GPT-4o and o3 models on long-context retrieval and multi-hop long-context reasoning\footnote{https://openai.com/index/gpt-4-1/}. These benchmarks indicate that GPT-4.1 is particularly reliable at retrieving, tracking, and integrating information scattered across extended contexts. Such characteristics align directly with the needs of Stage~2, where the model must synthesise behavioural descriptions across an entire video without speculative inference. For these reasons, we selected GPT-4.1 as the backbone for Stage~2 rather than reusing the Stage~1 model or employing a specialised reasoning model.

Evaluation in Stage~2 followed the same holistic procedure used by SLPs when annotating parent–child interactions. All segments from the same video were provided together in a single prompt, allowing the model to review the full sequence of behaviours before making any judgement. Each segment was represented as a structured pair of \((\textit{parent}, \textit{child})\) observations across gaze, action, and vocalisation. Rather than assigning labels to segments independently, the model was required to integrate information across the entire video and identify which parts of the interaction corresponded to \textit{Strong}, \textit{Moderate}, or \textit{Poor} joint attention, mirroring the way SLPs derive their final ratings from the full interaction.

To assess how SLPs’ professional knowledge supports the alignment of MLLMs with expert judgement, we designed two prompting conditions: a \textit{zero-shot} condition without access to SLP examples, and a \textit{many-shot} condition that incorporated many SLP-annotated examples.
The model was then asked to judge \emph{all segments} of the held-out target video (grouped together to preserve contextual continuity), returning one label per segment in the prescribed format.

Drawing from our earlier interviews, we observed that SLPs typically arrive at a judgement through three steps: observing behavioural cues, reasoning about social coordination, and mapping these to categorical Judgements (e.g., strong, moderate, poor). We therefore designed the prompt structure to mirror this expert reasoning pipeline.

In the \textit{zero-shot} condition, the prompt contained only task instructions, and the model produced judgements directly from the observational descriptions.  

\noindent\textbf{\textit{Zero-shot prompt}}:
\begin{quote}\ttfamily
\small
You are a speech-language pathologist. Joint attention in a child refers to the ability to share attention with another person--typically the parent--by coordinating behaviours such as actions, vocalisations (e.g., speaking or making sounds), or gaze (looking at shared objects or people).\\
Please evaluate the quality of the child's joint attention in each segment below based on their behaviours. Respond using the following format:\\
Segment 1: [Strong/Moderate/Poor] \\
Segment 2: ...
\end{quote}

In the \textit{many-shot} condition, we adopted a leave-one-video-out strategy~\cite{zhao2024learning}: the held-out video served as the evaluation set, while annotated examples from all remaining videos-each consisting of \((\textit{parent}, \textit{child})\) observations paired with a gold label-were embedded in the prompt as demonstrations. This design ensured that the model had access to the full range of SLP-provided knowledge from other videos when making judgements on the target segments. The procedure was repeated across all videos, so that every video was evaluated once as the hold-out, and the union of predictions from all folds yielded the complete set of model-generated judgement labels.

Below is a simplified illustration of the \textit{many-shot} prompting structure used in our experiments. 

\noindent\textbf{\textit{Many-shot examples}}:
\begin{quote}\ttfamily
\small
Below are labelled examples from other videos by real speech-language therapists. Each example shows a structured observational description and its judgement label.\\
...\\
Example 124\\
Parent:\\
- Action: The parent sat at the table with their hands clasped.\\
- Vocalisation: The parent said ``Maybe, possibly''.\\
- Gaze: The parent looked at the child and the game board.\\
Child:\\
- Action: The child placed a white game piece onto the game board.\\
- Vocalisation: None\\
- Gaze: The child looked at the game board and the pieces.\\
Judgement: \textit{Moderate}\\
...\\
Example 178
Parent:\\
- Action: The parent sat at the table with her hands resting on it.\\
- Vocalisation: The parent said ``Tricky. Ooh, you switched it. I like this.''\\
- Gaze: The parent looked at the child and the game board.\\
Child:\\
- Action: The child moved a white game piece across the board.\\
- Vocalisation: None\\
- Gaze: The child looked down at the game board.\\
Judgement: \textit{Strong}\\  
...\\
Example 273  
Parent:\\
- Action: The parent sat at the table with hands clasped.\\
- Vocalisation: The parent said ``Are you getting the trains ready?''\\
- Gaze: The parent looked at the child.\\
Child:  
- Action: The child pushed wooden train track pieces together on the table.\\
- Vocalisation: The child said ``Train''.\\
- Gaze: The child looked down at the train track pieces.\\
Judgement: \textit{Poor}
\end{quote}

For both the zero-shot and many-shot settings, we evaluated the model using three decoding temperatures (0, 0.3, and 0.5). Because the judgment task requires stable and reliable decisions rather than highly diverse generations, we intentionally restricted the temperature range to relatively low values. To account for the stochasticity introduced at temperatures above zero, we conducted three independent runs for each non-deterministic setting (0.3 and 0.5). Temperature 0 is deterministic and therefore required only a single run.

\subsubsection{Evaluation metrics}
To assess overall alignment, we compared model-generated labels with the \emph{agreement} label for each segment (i.e., the consensus label retained in the agreement dataset, requiring at least two SLPs to agree). We report \textit{accuracy}; \textit{macro-precision}, \textit{macro-recall}, and \textit{macro-F1} (unweighted averages over \textit{Strong}, \textit{Moderate}, \textit{Poor} to mitigate class imbalance); and \textit{Cohen’s~$\kappa$} to quantify agreement beyond chance. All metrics are computed at the segment level across the full agreement dataset. In addition to macro-level scores, we also report per-class precision, recall, and F1 to characterise performance on each judgement category.

\subsubsection{Result: Many-shot Prompting Strategy Yields Better Alignment}

\begin{figure*}[ht]
    \centering
    \includegraphics[width=\textwidth]{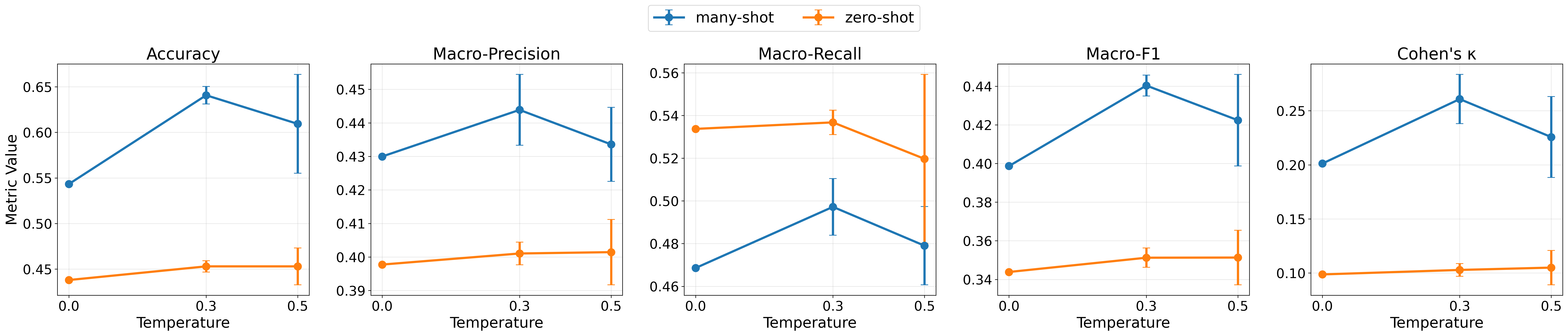}
    \caption{Effect of decoding temperature on model reliability for zero-shot and many-shot prompting.
    Each subplot shows mean performance with standard-deviation error bars across temperatures 0, 0.3, and 0.5 for accuracy, macro-precision, macro-recall, macro-F1, and Cohen’s~$\kappa$.
    Across all metrics, temperature~0.3 provides the most reliable overall performance, while temperature~0 underperforms and temperature~0.5 exhibits reduced stability due to increased sampling randomness.
    Many-shot prompting benefits more from moderate stochasticity than zero-shot prompting, which remains comparatively less sensitive to temperature changes.
    These results support the choice of temperature~0.3 for subsequent analyses.}
    \Description{Five side-by-side line plots compare accuracy, macro-precision, macro-recall, macro-F1, and Cohen’s kappa across temperatures 0, 0.3, and 0.5 for zero-shot and many-shot prompting. Temperature 0.3 yields the most stable and consistent performance, temperature 0 performs worse overall, and temperature 0.5 shows higher variance. Many-shot prompting is more affected by temperature changes than zero-shot prompting.}
        \label{fig:temperature}
\end{figure*}

\paragraph{Temperature effects.}
Figure~\ref{fig:temperature} shows clear temperature-dependent differences in model reliability across both prompting modes. Temperature~0.3 consistently produced the strongest and most stable performance: in the many-shot condition it achieved the highest accuracy ($0.64 \pm 0.01$), macro-F1 ($0.44 \pm 0.01$), and Cohen’s~$\kappa$ ($0.26 \pm 0.02$), while zero-shot prompting likewise reached its best scores at this setting (macro-F1 $= 0.351 \pm 0.005$, $\kappa = 0.103 \pm 0.006$). In contrast, temperature~0 underperformed in both modes (e.g., many-shot macro-F1 $= 0.40$, zero-shot macro-F1 $= 0.344$). Temperature~0.5 yielded slightly lower means and substantially higher variance; for example, in many-shot prompting accuracy dropped to $0.61 \pm 0.05$ and macro-F1 to $0.42 \pm 0.02$, suggesting that excessive sampling randomness reduces stability. Overall, temperature~0.3 offers the best balance of accuracy and robustness, and we therefore adopt it as the default for subsequent analyses.

\begin{figure*}[ht]
    \centering
    \includegraphics[width=\textwidth]{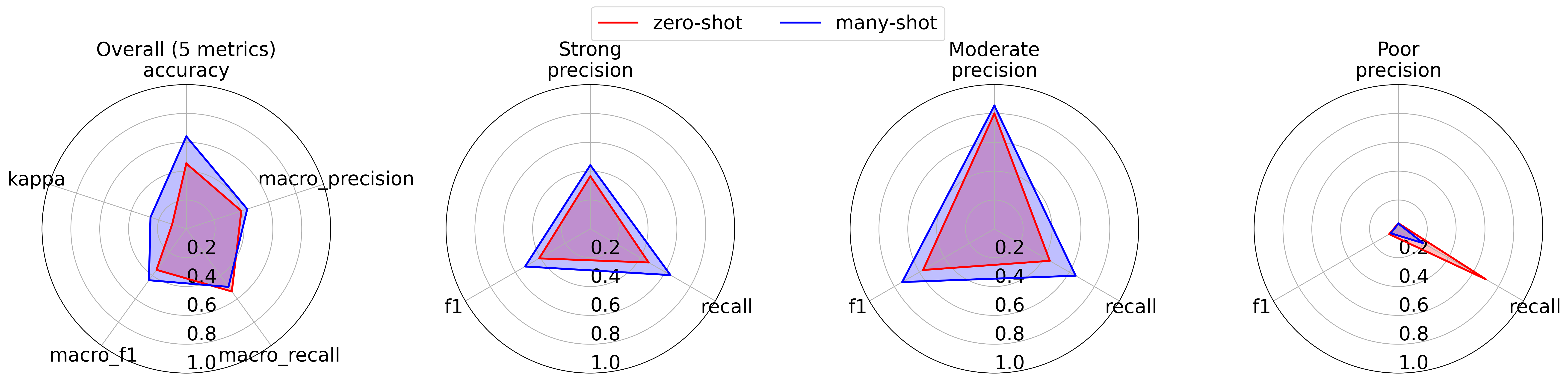}
    \caption{Radar plots comparing zero-shot (red) and many-shot (blue) prompting across overall and per-class evaluation metrics. many-shot consistently improves accuracy, macro-F1, and Cohen’s~$\kappa$ in the overall condition, while also boosting performance on \textit{Strong} and \textit{Moderate} categories. Performance on the \textit{Poor} category remains weak under both conditions, reflecting its limited representation in the dataset.}
    \Description{The radar plot compares the performance of zero-shot (red) and many-shot (blue) prompting, showing that many-shot prompting consistently improves overall metrics and performance for the 'Strong' and 'Moderate' categories, while performance on the 'Poor' category remains low due to its limited dataset representation.}
    \label{fig:model}
\end{figure*}

\paragraph{Overall performance}
Many-shot prompting achieved stronger alignment with expert consensus than zero-shot prompting (see Figure~\ref{fig:model}). Accuracy improved from 0.45 to 0.64, and Cohen’s~$\kappa$ increased from 0.10 to 0.26. While Cohen’s~$\kappa$ shows an improvement, it still remains within the slight-agreement range. Macro-precision also rose modestly (0.40 → 0.44), indicating fewer false positives, while macro-recall decreased slightly (0.54 → 0.50), suggesting a more conservative stance that misses some expert-labelled cases. Despite this trade-off, the overall macro-F1 increased from 0.35 to 0.44, representing a clear improvement in balanced performance.

\paragraph{Per-class performance.}
At the category level, many-shot prompting yields improvements for \textit{Strong} and \textit{Moderate} cases but not for \textit{Poor}.
For \textit{Strong}, recall increases from 0.47 to 0.64 and F1 improves from 0.41 to 0.52, indicating more reliable detection of clear joint attention.
For \textit{Moderate}, which dominates the dataset, both precision (0.80→0.85) and recall (0.44→0.65) increase, producing the largest F1 gain (0.57→0.74).
By contrast, \textit{Poor} segments remain highly challenging: recall drops sharply (0.70→0.20) and F1 decreases (0.07→0.06).
This reflects two issues. First, zero-shot prompting substantially over-predicts \textit{Poor} cases, producing high recall but extremely low precision (0.04). Many-shot prompting suppresses this over-prediction, yielding more conservative outputs but still very low precision (0.04→0.04).
Second, the dataset distribution is highly imbalanced: \textit{Moderate} accounts for 469 segments (73.5\%), \textit{Strong} for 136 (21.3\%), and \textit{Poor} for only 10 examples (1.6\%).
Such scarcity makes it difficult for the model to learn or validate reliable patterns for low-fidelity \textit{Poor} behaviours, regardless of prompting method.

%% file: sections/7_discussion.tex
\section{Discussion}

\subsection{Greater Alignment at the Level of Behavioural Observations}
SLPs showed strong alignment with one another in behavioural observation, consistently pointing to \ga{}, \ac{}, and \vo{} as the central dimensions for describing \ja{}. Our structured prompting method leveraged this shared consensus and achieved similarly high alignment. MLLMs could reliably capture and reproduce these behavioural cues. Because experts already converge on these descriptors, observation proved to be a tractable entry point for human–AI alignment.

This tractability reflects a broader phenomenon. When the domain involves low-level, directly perceivable behaviours, experts tend to describe events in highly similar ways. In education~\cite{xing2024survey}, teachers reliably agree on whether a child raised a hand or spoke out loud. In UX evaluation~\cite{vermeeren2010user}, experts consistently identify pauses, clicks, and hovers. In creative work critiques~\cite{alabood2023systematic}, reviewers often converge on surface-level cues such as colour saturation or bold outlines. These shared descriptors create a stable grounding that multimodal models can be scaffolded to replicate. As a result, observation-level alignment can be achieved before attempting any higher-level judging.

This suggests that systems for analysing \pci{} should prioritise clear and structured observation as their first layer of alignment. By guiding MLLMs to “see” behaviours through common expert descriptors, we can produce stable and interpretable outputs that form a reliable substrate for subsequent judging tasks.

\textbf{\textit{Design Takeaway \#1: Provide Perceptual Assistance With Transparent and Interpretable Cues}}.
MLLMs can meaningfully support early-stage observation. In our study, the model achieved 81\% accuracy in identifying low-level behavioural cues, showing strong alignment with SLPs in detecting gaze shifts, object-directed actions, and vocalisations. However, these machine-generated observations cannot replace human observation, especially in a clinically serious and nuanced behavioural assessment task. Even with relatively strong cue-level performance, SLPs should not simply trust model outputs. They must actively review, verify, and integrate these cues into their own observation workflow.

These cues nevertheless remain valuable. They are precisely the behaviours that SLPs spend substantial time documenting during and after sessions, both for clinical records and to support later judgement. SLP2 specifically noted that gathering concrete behavioural examples is a time-consuming but essential part of subsequent assessment. By surfacing structured and timestamped observations, MLLMs can reduce documentation burden and provide a consistent baseline for longitudinal tracking, provided that SLPs remain the primary interpreters.

This highlights a concrete design direction. Systems should help SLPs use model-extracted cues efficiently and reflectively. Interfaces should allow interactive editing of cue descriptions, timestamp correction, merging or splitting of events, and clinician-authored annotations. These features can transform model detections into editable professional artefacts. Transparent visualisations, such as layered timelines, collapsible cue categories, uncertainty indicators, and explicit evidence, can further support rapid verification while preventing over-reliance on automation. Through these mechanisms, perceptual assistance augments rather than replaces human expertise and embeds MLLM-derived cues into clinical judgement in a controlled and interpretable manner.

\subsection{Partial Alignment at the Level of Judgement}

Alignment at the judgement stage was modest, reflecting the difficulty of capturing interpretive nuances. Many-shot prompting consistently outperformed zero-shot prompting, suggesting that examples grounded in SLP expertise help the model approximate expert judging more closely. These gains were incremental, however, indicating that judgement-level alignment is shaped not only by prompting strategies but also by natural variation in expert perspectives.

Judgement in \pci{} analysis is inherently subjective and context dependent. Different SLPs weight behavioural cues differently, and their interpretations can diverge even when they observe the same interaction. Similar variation also appears in other expert domains. Teachers may interpret a raised hand as engagement, whereas others see it as disruption; UX evaluators may view a pause as confusion, while others treat it as normal exploration; and reviewers in creative critique often make contrasting assessments of the same work. These examples illustrate that judgement alignment is fundamentally different from observation alignment.

In our dataset, heterogeneous annotations made it difficult to derive a single, coherent judgement space. We also observed pockets of higher agreement among certain experts, such as SLP1 and SLP2 ($\alpha = 0.31$). When data are limited and expert perspectives vary widely, such internally consistent subgroups point to a possible direction for future work: examining whether alignment strategies that draw on individual experts or small groups could offer more stable early-stage models.

\textbf{\textit{Design Takeaway \#2: Calibrate AI Judgement to Task Needs Through Expert-Aligned Adaptation.}}
Given the heterogeneous annotation landscape, a practical direction for HCI systems is to support personalised, expert-aligned adaptation. For SLP-facing tools, tailoring the AI to an individual clinician’s own judgement tendencies may be particularly useful, as their criteria for a given task tend to remain relatively stable over time. For parent-facing tools, however, a consensus-informed representation derived from multiple experts may provide clearer and more balanced guidance. Developing such a system responsibly would require substantially larger and more diverse datasets, including naturalistic parent–child interactions and broader SLP annotations, to ensure that any consensus model is reliable, generalisable, and appropriate for use in parent-oriented contexts.

At the presentation level, SLPs also highlighted the importance of conveying joint attention as an overall pattern rather than a sequence of isolated segments. SLP3 suggested a colour-bar overview where each segment’s label is encoded as a colour across the entire timeline. Such visual summaries provide both SLPs and parents with an immediate, intuitive understanding of interaction quality and help them focus efficiently on segments that may require further attention.

Across these use cases, AI outputs should be treated as supportive references rather than definitive judgements. From a reliability standpoint, the final assessment should still be made by an SLP. Judgement in \pci{} is only one part of an SLP’s broader role; clinicians also provide therapy, coach parents, interact with children, and tailor guidance to highly specific family contexts. These relational and instructional responsibilities cannot be automated. AI systems can, however, enhance clinical efficiency by surfacing behavioural patterns, organising information, and providing structured summaries that clinicians can interpret and incorporate into therapeutic practice. Future work should investigate how personalised expert aligned models and consensus informed models might be integrated into broader clinical workflows and how these two modes could support different stakeholders within real developmental assessment and intervention pathways.

\subsection{Limitations}

\paragraph{Sample size}  
Our study involved only three SLPs, which necessarily limits the robustness and generalisability of the findings. 
While their insights were rich and directly informed our two-stage pipeline design, the small $N=3$ means the patterns we report should be read as illustrative rather than definitive. 
Future work will need to engage a larger and more diverse pool of practitioners to examine whether the alignment and misalignment patterns we observed hold across settings and populations.

\paragraph{Dataset characteristics}
The dataset of 25 YouTube videos was skewed toward short, likely neurotypical interactions, introducing bias and constraining generalisability.
Our contribution is exploratory, and we used demonstration-based recordings as a pragmatic choice given the difficulty of collecting authentic, neurodiverse, and naturally occurring \pci{} data.
However, this bias led to very few segments labelled as \textit{Poor}, meaning our analysis primarily reflects alignment in the \textit{Strong} category and provides only partial evidence of model–expert agreement overall.
Notably, one SLP emphasised that detecting \textit{Poor} joint attention is often clinically more critical, and suggested that recordings featuring neurodiverse children (e.g., autism-focused datasets) may not only be more representative of practice but also easier to annotate consistently.
Expanding future datasets to include longer interactions, neurodiverse populations~\cite{benton2014diversity}, and recordings from real-world home or school settings will be essential for improving ecological validity.

\paragraph{Human Editing}
A methodological limitation of our approach is that the input to the Stage~2 judge process was based on manually corrected descriptions rather than the raw outputs produced by the MLLM in Stage~1. Although fewer than 15\% of all entries required correction and each modification followed a minimal-editing principle that preserved the subject--verb--object structure, these corrections removed factual errors that would otherwise propagate into Stage~2. Our intention was to isolate the effect of the judge process itself, but this design choice means that the Stage~2 results likely represent an upper bound on performance in a fully end-to-end setting. Future work will evaluate the judge process using uncorrected model outputs to obtain a more conservative estimate of system behaviour.

\paragraph{Positioning and Scope}
It is important to clarify the scope of this work. 
We do not propose or validate any clinical framework for diagnosis or intervention, nor do we claim clinical applicability. 
Rather, we present an exploratory, case-based attempt to align model outputs with expert perspectives in parent–child interaction analysis. 
Our aim is to examine where alignment appears feasible at the level of observable behaviour, and where it becomes more challenging at the level of interpretive judgement, thereby surfacing design questions for future HCI systems.
In this sense, the system is a probe rather than a contribution in itself: it is used to study alignment patterns, not to introduce a novel prompting technique or clinical tool. 
Any move toward practical use would require larger and more varied datasets, broader expert participation, and rigorous validation, which are beyond the scope of this study.

%% file: sections/8_conclusion.tex
\section{Conclusion}
This paper presented an exploratory study of alignment in the context of parent–child interaction analysis. By comparing three SLP’ perspectives on \ja{} with MLLM outputs, we identified a clear contrast: observation-level alignment was relatively robust, while judgement-level alignment remained elusive due to differences among experts themselves. These findings suggest that MLLMs can serve as reliable observers when scaffolded by shared behavioural cues, and they highlight how future HCI systems may draw value from both individual or subgroup SLP judgements and areas of broader consensus across the expert cohort.

Our contribution should be understood as a case-based investigation of alignment rather than a validated framework. The study highlights design opportunities for HCI and AI alignment research more broadly, where systems may act as collaborators that surface observations and alternative interpretations. 

%% file: sections/appendix.tex
\clearpage 
\appendix
\section{Video Dataset Appendix}
\label{app:video-list}
\begin{center}
\small
\begin{tabular}{p{14cm}}
\toprule
\textbf{Youtube Videos (with link)} \\
\midrule

\multicolumn{1}{c}{\textbf{Behavioral Guidance and Skill Modeling}} \\
\midrule
Video 2: \href{https://www.youtube.com/watch?v=Wxd3BHF8bKU}{Parent Models All PRIDE Skills \& Active Ignoring- Parent-Child Interaction Therapy (PCIT)} \\

Video 3: \href{https://youtu.be/YUkujhg6j6w?si=Yfcb56YTqcmS5w3l}{Therapist Coaches Big Ignore- Parent-Child Interaction Therapy (PCIT) for Older Children} \\

Video 5: \href{https://youtu.be/zeaAX596T2g?si=Bd2roO6OhpQMOupY}{Parent Uses -Big Ignore- Technique- Parent-Child Interaction Therapy (PCIT) for Older Children} \\

Video 6: \href{https://youtu.be/45GSMS6KrLA?si=7yoCq2xZTANDuv_1}{Parent Models Skill to AVOID, Commands-- Parent-Child Interaction Therapy (PCIT)} \\

Video 8: \href{https://youtu.be/YFK9G5d7r5g?si=2pw4sh-V6-x_p4xV}{Parent Models Skill to AVOID, Negative Talk-- Parent-Child Interaction Therapy (PCIT)} \\

Video 9: \href{https://youtu.be/9jiwyrz4gTQ?si=pTKsajEeuyO6-I6p}{Parent Models PRIDE Skill -P for Praise-- Parent-Child Interaction Therapy (PCIT)} \\

Video 11: \href{https://youtu.be/6wvrzzoxRrE?si=KkwEMTEoCeWqcHKG}{Parent Models PRIDE Skill -R for Reflection-- Parent-Child Interaction Therapy (PCIT)} \\

Video 12: \href{https://youtu.be/kkDj4pzzP9Q?si=Cjhy5fO8o_E30PzX}{Parent Models PRIDE Skill -I for Imitation-- Parent-Child Interaction Therapy (PCIT)} \\

Video 13: \href{https://youtu.be/XE43uu3OSxg?si=sDSqQFh8h9TdG9Zf}{Parent Models PRIDE Skill -D for Description-- Parent-Child Interaction Therapy (PCIT)} \\

Video 14: \href{https://youtu.be/2xmTGLsiS08?si=znMGMHN2BXFZVViS}{Parent Models PRIDE Skill -E for Enthusiasm-Enjoyment-- Parent-Child Interaction Therapy (PCIT)} \\

Video 16: \href{https://youtu.be/quveCprzfxE?si=k8V3H-cL9kjGe0OM}{Parent Models Differential Attention-Active Ignoring- Parent-Child Interaction Therapy (PCIT)} \\

\midrule
\multicolumn{1}{c}{\textbf{Language and Cognitive Development}} \\
\midrule

Video 17: \href{https://youtu.be/lhHkJ3InQOE?si=kXegR6ey-36BG_kL}{A Typical 10-month-old on Piaget's A-not-B task} \\

Video 18: \href{https://youtu.be/tXZau5VIIvU?si=C-A-Ct2PQIYt5hJf}{A Typical 3-year-old Sorting cards} \\

Video 19: \href{https://youtu.be/gnArvcWaH6I?si=Sp1v9FXAtTe-ipUf}{A typical child on Piaget's conservation tasks} \\

Video 20: \href{https://youtu.be/-UIrOZkYmO4?si=58fvHAt0zuyNKlNW}{ABA Sample Session (cards and chase)} \\

Video 21: \href{https://youtu.be/RzI6Ar5mu2Q?si=UN6IIxtpcbeRiRhH}{ABA Therapy - Learning about Animals} \\

Video 22: \href{https://youtu.be/V9YDDpo9LWg?si=Dibk5ufFFFyhyZjs}{ABA Therapy- Daniel - Communication} \\

Video 23: \href{https://youtu.be/VV9pOkmnKCY?si=_GYNWQ3CFUXadBQG}{Encouraging Language Development- A Positive Parent-Child Interaction} \\

Video 24: \href{https://www.youtube.com/watch?v=rVqJacvywAQ}{Piaget - Object permanence failure (Sensorimotor Stage)} \\

Video 25: \href{https://youtu.be/4jW668F7HdA?si=xrzumAzGAwa82Rte}{Piaget - The A Not B Error (Sensorimotor Stage)} \\

\midrule
\multicolumn{1}{c}{\textbf{Daily Life Skills and Interaction}} \\
\midrule
Video 1: \href{https://youtu.be/7DVdgwbGEpU?si=DHrTC1E02BNkiGPI}{Parent Models Skill to AVOID, Questions-- Parent-Child Interaction Therapy (PCIT)} \\

Video 4: \href{https://youtu.be/ds7wXB0nlz4?si=X1PEHzv9dokjZhOq}{Parent Models How to End Special Playtime- Parent-Child Interaction Therapy (PCIT)} \\

Video 7: \href{https://youtu.be/aL2pV6DMgLY?si=FlUeDT3wtyY-r9bB}{Parent \& Child Check Sticker Chart- Parent-Child Interaction Therapy (PCIT) for Older Children} \\

Video 10: \href{https://youtu.be/N3wAPLXd7I0?si=UzP8TiWjoWYNg2jq}{Parent Meets CDI Goal Criteria- Parent-Child Interaction Therapy (PCIT) for Older Children} \\

Video 15: \href{https://youtu.be/pYVU4fcH8Oc?si=GTchQKgECvVB-D4I}{Child Complies with Command in PDI- Parent-Child Interaction Therapy (PCIT) for Older Children} \\

\bottomrule
\end{tabular}
\end{center}

%% file: main.bib
@article{sunderajan_speech_2019,
	title = {Speech and language delay in children: {Prevalence} and risk factors},
	volume = {8},
	issn = {2249-4863},
	shorttitle = {Speech and language delay in children},
	url = {https://www.ncbi.nlm.nih.gov/pmc/articles/PMC6559061/},
	doi = {10.4103/jfmpc.jfmpc_162_19},
	number = {5},
	urldate = {2025-05-14},
	journal = {Journal of Family Medicine and Primary Care},
	author = {Sunderajan, Trisha and Kanhere, Sujata V.},
	month = may,
	year = {2019},
	pmid = {31198730},
	pmcid = {PMC6559061},
	pages = {1642--1646},
}

@article{robinson2017cdc,
  title={CDC grand rounds: Addressing health disparities in early childhood},
  author={Robinson, Lara R},
  journal={MMWR. Morbidity and mortality weekly report},
  volume={66},
  year={2017}
}

@article{tomasello1986joint,
  title={Joint attention and early language},
  author={Tomasello, Michael and Farrar, Michael Jeffrey},
  journal={Child development},
  pages={1454--1463},
  year={1986},
  publisher={JSTOR}
}

@article{lu2024gpt,
  title={From gpt-4 to gemini and beyond: Assessing the landscape of mllms on generalizability, trustworthiness and causality through four modalities},
  author={Lu, Chaochao and Qian, Chen and Zheng, Guodong and Fan, Hongxing and Gao, Hongzhi and Zhang, Jie and Shao, Jing and Deng, Jingyi and Fu, Jinlan and Huang, Kexin and others},
  journal={arXiv preprint arXiv:2401.15071},
  year={2024}
}

@article{team2024gemini,
  title={Gemini 1.5: Unlocking multimodal understanding across millions of tokens of context},
  author={Team, Gemini and Georgiev, Petko and Lei, Ving Ian and Burnell, Ryan and Bai, Libin and Gulati, Anmol and Tanzer, Garrett and Vincent, Damien and Pan, Zhufeng and Wang, Shibo and others},
  journal={arXiv preprint arXiv:2403.05530},
  year={2024}
}

@inproceedings{bain2023whisperx,
  title={Whisperx: Time-accurate speech transcription of long-form audio},
  author={Bain, Max and Huh, Jaesung and Han, Tengda and Zisserman, Andrew},
  booktitle = {Proceedings of the International Speech Communication Association (INTERSPEECH)},
  year      = {2023},
  pages     = {4489--4493}
}

@inproceedings{song2016talklime,
  title={TalkLIME: mobile system intervention to improve parent-child interaction for children with language delay},
  author={Song, Seokwoo and Kim, Seungho and Kim, John and Park, Wonjeong and Yim, Dongsun},
  booktitle={Proceedings of the 2016 ACM International Joint Conference on Pervasive and Ubiquitous Computing},
  pages={304--315},
  year={2016}
}

@inproceedings{hwang2014talkbetter,
  title={TalkBetter: family-driven mobile intervention care for children with language delay},
  author={Hwang, Inseok and Yoo, Chungkuk and Hwang, Chanyou and Yim, Dongsun and Lee, Youngki and Min, Chulhong and Kim, John and Song, Junehwa},
  booktitle={Proceedings of the 17th ACM conference on Computer supported cooperative work \& social computing},
  pages={1283--1296},
  year={2014}
}

@online{dir_floortime,
  title={DIR Floortime},
  author={{The Interdisciplinary Council on Development and Learning}},
  year= {2025},
  url={https://www.icdl.com/dir/floortime},
}

@book{pepper2004talk,
  title={It Takes Two to Talk: A Practical Guide for Parents of Children with Language Delays},
  author={Jan Pepper and Elaine Weitzman},
  year={2004},
  publisher={Hanen Centre},
  note={Shows parents how to help their child communicate and learn language during everyday activities.},
}

@book{sussman1999words,
  title={More Than Words: A Guide to Helping Parents Promote Communication and Social Skills in Children with Autism Spectrum Disorder},
  author={Fern Sussman},
  year={1999},
  publisher={Hanen Centre},
  note={Step by step guide for parents of preschool children with autism spectrum disorder and other social communication difficulties.},
}

@article{o2005barriers,
  title={Barriers to accessing rural paediatric speech pathology services: Health care consumers’ perspectives},
  author={O'Callaghan, Anna M and McAllister, Lindy and Wilson, Linda},
  journal={Australian Journal of Rural Health},
  volume={13},
  number={3},
  pages={162--171},
  year={2005},
  publisher={Wiley Online Library}
}

@inproceedings{chan2017wakey,
  title={WAKEY: assisting parent-child communication for better morning routines},
  author={Chan, Meng-Ying and Lin, Yi-Hsuan and Lin, Long-Fei and Lin, Ting-Wei and Hsu, Wei-Che and Chang, Chia-yu and Liu, Rui and Chang, Ko-Yu and Lin, Min-hua and Hsu, Jane Yung-jen},
  booktitle={Proceedings of the 2017 ACM Conference on Computer Supported Cooperative Work and Social Computing},
  pages={2287--2299},
  year={2017}
}

@inproceedings{kwon2022captivate,
  title={Captivate! contextual language guidance for parent--child interaction},
  author={Kwon, Taeahn and Jeong, Minkyung and Ko, Eon-Suk and Lee, Youngki},
  booktitle={Proceedings of the 2022 CHI Conference on Human Factors in Computing Systems},
  pages={1--17},
  year={2022}
}

@inproceedings{zheng2024soap,
  title={SOAP. AI: A Collaborative Tool for Documenting Human Behavior in Videos through Multimodal Generative AI},
  author={Zheng, Qingxiao and Rabbani, Parisa and Lin, Yu-Rou and Mansour, Daan and Huang, Yun},
  booktitle={Companion Publication of the 2024 Conference on Computer-Supported Cooperative Work and Social Computing},
  pages={87--90},
  year={2024}
}

@article{mclaughlin2011speech,
  title={Speech and language delay in children},
  author={McLaughlin, Maura R},
  journal={American family physician},
  volume={83},
  number={10},
  pages={1183--1188},
  year={2011}
}

@article{sunderajan2019speech,
  title={Speech and language delay in children: Prevalence and risk factors},
  author={Sunderajan, Trisha and Kanhere, Sujata V},
  journal={Journal of family medicine and primary care},
  volume={8},
  number={5},
  pages={1642--1646},
  year={2019},
  publisher={Medknow}
}

@article{mundy2007individual,
  title={Individual differences and the development of joint attention in infancy},
  author={Mundy, Peter and Block, Jessica and Delgado, Christine and Pomares, Yuly and Van Hecke, Amy Vaughan and Parlade, Meaghan Venezia},
  journal={Child development},
  volume={78},
  number={3},
  pages={938--954},
  year={2007},
  publisher={Wiley Online Library}
}

@inproceedings{lewis2025exploring,
  title={Exploring AI-Based Support in Speech-Language Pathology for Culturally and Linguistically Diverse Children},
  author={Lewis, Aaleyah and Dangol, Aayushi and Suh, Hyewon and Olszewski, Abbie and Fogarty, James and Kientz, Julie A},
  booktitle={CHI Conference on Human Factors in Computing Systems (CHI’25)},
  year={2025},
  organization={ACM New York, NY, USA}
}

@inproceedings{dangol2025want,
  title={“I Want to Think Like an SLP”: A Design Exploration of AI-Supported Home Practice in Speech Therapy},
  author={Dangol, Aayushi and Lewis, Aaleyah and Suh, Hyewon and Hong, Xuesi and Meadan, Hedda and Fogarty, James and Kientz, Julie A},
  booktitle={Proceedings of the 2025 CHI Conference on Human Factors in Computing Systems},
  pages={1--21},
  year={2025}
}

@article{masse2018taking,
  title={Taking PRIDE in your home: Implementing home-based Parent--Child Interaction Therapy (PCIT) with fidelity},
  author={Masse, Joshua J and Quetsch, Lauren Borduin and McNeil, Cheryl B},
  journal={Handbook of parent-child interaction therapy: Innovations and applications for research and practice},
  pages={161--181},
  year={2018},
  publisher={Springer}
}

@article{woodfield2021time,
  title={Time-out with young children: a parent-child interaction therapy (PCIT) practitioner review},
  author={Woodfield, Melanie J and Brodd, Irene and Hetrick, Sarah E},
  journal={International journal of environmental research and public health},
  volume={19},
  number={1},
  pages={145},
  year={2021},
  publisher={MDPI}
}

@inproceedings{choi2025aacess,
  title={AACessTalk: Fostering Communication between Minimally Verbal Autistic Children and Parents with Contextual Guidance and Card Recommendation},
  author={Choi, Dasom and Park, SoHyun and Lee, Kyungah and Hong, Hwajung and Kim, Young-Ho},
  booktitle={Proceedings of the 2025 CHI Conference on Human Factors in Computing Systems},
  pages={1--25},
  year={2025}
}

@article{brock2014statewide,
  title={Statewide assessment of professional development needs related to educating students with autism spectrum disorder},
  author={Brock, Matthew E and Huber, Heartley B and Carter, Erik W and Juarez, A Pablo and Warren, Zachary E},
  journal={Focus on Autism and Other Developmental Disabilities},
  volume={29},
  number={2},
  pages={67--79},
  year={2014},
  publisher={Sage Publications Sage CA: Los Angeles, CA}
}

@article{finke2009all,
  title={“All children can and should have the opportunity to learn”: General education teachers' perspectives on including children with autism spectrum disorder who require AAC},
  author={Finke, Erinn H and Finke, Erinn H and McNaughton, David B and Drager, Kathryn DR},
  journal={Augmentative and Alternative Communication},
  volume={25},
  number={2},
  pages={110--122},
  year={2009},
  publisher={Taylor \& Francis}
}

@article{gadberry2011survey,
  title={A survey of the use of aided augmentative and alternative communication during music therapy sessions with persons with autism spectrum disorders.},
  author={Gadberry, Anita L},
  journal={Journal of music therapy},
  volume={48},
  number={1},
  year={2011}
}

@article{ganz2013impacts,
  title={Impacts of a PECS instructional coaching intervention on practitioners and children with autism},
  author={Ganz, Jennifer B and Goodwyn, Fara D and Boles, Margot M and Hong, Ee Rea and Rispoli, Mandy J and Lund, Emily M and Kite, Elizabeth},
  journal={Augmentative and Alternative Communication},
  volume={29},
  number={3},
  pages={210--221},
  year={2013},
  publisher={Taylor \& Francis}
}

@article{gandhi2023multimodal,
  title={Multimodal sentiment analysis: A systematic review of history, datasets, multimodal fusion methods, applications, challenges and future directions},
  author={Gandhi, Ankita and Adhvaryu, Kinjal and Poria, Soujanya and Cambria, Erik and Hussain, Amir},
  journal={Information Fusion},
  volume={91},
  pages={424--444},
  year={2023},
  publisher={Elsevier}
}

@inproceedings{jain2024vcoder,
  title={Vcoder: Versatile vision encoders for multimodal large language models},
  author={Jain, Jitesh and Yang, Jianwei and Shi, Humphrey},
  booktitle={Proceedings of the IEEE/CVF Conference on Computer Vision and Pattern Recognition},
  pages={27992--28002},
  year={2024}
}

@inproceedings{dos2023composite,
  title={Composite AI for behavior analysis in social interactions},
  author={Dos Santos Melicio, Bruno Carlos and Xiang, Linyun and Dillon, Emily and Soorya, Latha and Chetouani, Mohamed and Sarkany, Andras and Kun, Peter and Fenech, Kristian and Lorincz, Andras},
  booktitle={Companion Publication of the 25th International Conference on Multimodal Interaction},
  pages={389--397},
  year={2023}
}

@inproceedings{whitehead2024generative,
  title={The generative multimodal analysis (gma) methodology for studying socially shared regulation in collaborative learning},
  author={Whitehead, Ridwan and Nguyen, Andy and J{\"a}rvel{\"a}, Sanna},
  booktitle={The International Conference on Learning Analytics \& Knowledge (LAK24)},
  year={2024}
}

@online{straitstimes2024earlyintervention,
  author       = {{The Straits Times}},
  title        = {Long wait times for early intervention push parents of kids with developmental needs to private sector},
  year         = {2024},
  url          = {https://www.straitstimes.com/singapore/long-wait-times-for-early-intervention-push-parents-of-kids-with-developmental-needs-to-private-sector},
  note         = {Accessed: 2025-05-14},
  organization = {The Straits Times}
}

@article{lieneman2017parent,
  title={Parent--child interaction therapy: Current perspectives},
  author={Lieneman, Corey C and Brabson, Laurel A and Highlander, April and Wallace, Nancy M and McNeil, Cheryl B},
  journal={Psychology research and behavior management},
  pages={239--256},
  year={2017},
  publisher={Taylor \& Francis}
}

@inproceedings{fu2024video,
  title={Video-mme: The first-ever comprehensive evaluation benchmark of multi-modal llms in video analysis},
  author={Fu, Chaoyou and Dai, Yuhan and Luo, Yongdong and Li, Lei and Ren, Shuhuai and Zhang, Renrui and Wang, Zihan and Zhou, Chenyu and Shen, Yunhang and Zhang, Mengdan and others},
  booktitle={Proceedings of the Computer Vision and Pattern Recognition Conference},
  pages={24108--24118},
  year={2025}
}

@article{yuan2024designing,
  title={Designing Collaborative Technology for Intergenerational Social Play over Distance},
  author={Yuan, Ye and Jin, Qiao and Mills, Chelsea and Yarosh, Svetlana and Neustaedter, Carman},
  journal={Proceedings of the ACM on Human-Computer Interaction},
  volume={8},
  number={CSCW2},
  pages={1--26},
  year={2024},
  publisher={ACM New York, NY, USA}
}

@article{sun2024exploring,
  author = {Sun, Yuling and Chen, Jiaju and Yao, Bingsheng and Liu, Jiali and Wang, Dakuo and Ma, Xiaojuan and Lu, Yuxuan and Xu, Ying and He, Liang},
  title = {Exploring Parent's Needs for Children-Centered AI to Support Preschoolers' Interactive Storytelling and Reading Activities},
  journal={Proceedings of the ACM on Human-Computer Interaction},
  number={CSCW2},
  year = {2024},
}

@article{fiani2024exploring,
  title={Exploring the perspectives of social VR-aware non-parent adults and parents on children's use of social virtual reality},
  author={Fiani, Cristina and Saeghe, Pejman and McGill, Mark and Khamis, Mohamed},
  journal={Proceedings of the ACM on Human-Computer Interaction},
  volume={8},
  number={CSCW1},
  pages={1--25},
  year={2024},
  publisher={ACM New York, NY, USA}
}

@article{nikkhah2024family,
  title={Family Resilience in Care Coordination Technologies: Designing for Families as Adaptive Systems},
  author={Nikkhah, Sarah and Rode, Akash Uday and Kulkarni, Neha Keshav and Mittal, Priyanjali and Mueller, Emily L and Miller, Andrew D},
  journal={Proceedings of the ACM on Human-Computer Interaction},
  volume={8},
  number={CSCW2},
  pages={1--28},
  year={2024},
  publisher={ACM New York, NY, USA}
}

@article{currin2024opportunities,
  title={Opportunities and Challenges in Using Tangible, Teleoperated Voice Agents in Kid-Driven Moments in Play Among Families with Neurodivergent Children},
  author={Currin, Flannery Hope and Kilcoin, Cassidy and Peterman, Kerry and Rector, Kyle and Hourcade, Juan Pablo},
  journal={Proceedings of the ACM on human-computer interaction},
  volume={8},
  number={CSCW1},
  pages={1--25},
  year={2024},
  publisher={ACM New York, NY, USA}
}

@article{su2024hidden,
  title={The Hidden Burden: Encountering and Managing (Unintended) Stigma in Children with Serious Illnesses},
  author={Su, Zhaoyuan and Kamath, Sunil P and Tirakitsoontorn, Pornchai and Chen, Yunan},
  journal={Proceedings of the ACM on Human-Computer Interaction},
  volume={8},
  number={CSCW1},
  pages={1--35},
  year={2024},
  publisher={ACM New York, NY, USA}
}

@inproceedings{shi2025towards,
  title={Towards multimodal large-language models for parent-child interaction: A focus on joint attention},
  author={Shi, Weiyan and Le, Hai Viet and Choo, Kenny Tsu Wei},
  booktitle={Proceedings of the Extended Abstracts of the CHI Conference on Human Factors in Computing Systems},
  pages={1--6},
  year={2025}
}

@inproceedings{li2025asd,
  title={ASD-HI: A Parent-Child Interaction Dataset for Automated Assessment of Home Intervention},
  author={Li, Zhaohui and Akemoglu, Yusuf and Lyu, Jincheng and Zheng, Qingxiao and Xiong, Jinjun},
  booktitle={International Conference on Artificial Intelligence in Education},
  pages={48--62},
  year={2025},
  organization={Springer}
}

@inproceedings{zheng2025ai,
  title={AI-Enhanced Speech-Language Intervention Documentation: Opportunities and Design Goals},
  author={Zheng, Qingxiao and Choudhry, Abhinav and Liu, Zihan and Rabbani, Parisa and Hu, Yuting and Olszewski, Abbie and Huang, Yun and Xiong, Jinjun},
  booktitle={International Conference on Artificial Intelligence in Education},
  pages={132--140},
  year={2025},
  organization={Springer}
}

@article{xing2024survey,
  title={A survey on MLLMs in education: application and future directions},
  author={Xing, Weicheng and Zhu, Tianqing and Wang, Jenny and Liu, Bo},
  journal={Future Internet},
  year={2024},
  publisher={MDPI}
}

@inproceedings{vermeeren2010user,
  title={User experience evaluation methods: current state and development needs},
  author={Vermeeren, Arnold POS and Law, Effie Lai-Chong and Roto, Virpi and Obrist, Marianna and Hoonhout, Jettie and V{\"a}{\"a}n{\"a}nen-Vainio-Mattila, Kaisa},
  booktitle={Proceedings of the 6th Nordic conference on human-computer interaction: Extending boundaries},
  pages={521--530},
  year={2010}
}

@article{alabood2023systematic,
  title={A systematic literature review of the Design Critique method},
  author={Alabood, Lorans and Aminolroaya, Zahra and Yim, Dianna and Addam, Omar and Maurer, Frank},
  journal={Information and Software Technology},
  volume={153},
  pages={107081},
  year={2023},
  publisher={Elsevier}
}

@inproceedings{benton2014diversity,
  title={Diversity for design: a framework for involving neurodiverse children in the technology design process},
  author={Benton, Laura and Vasalou, Asimina and Khaled, Rilla and Johnson, Hilary and Gooch, Daniel},
  booktitle={Proceedings of the SIGCHI conference on Human Factors in Computing Systems},
  pages={3747--3756},
  year={2014}
}

@article{zhao2024learning,
  title={Learning domain invariant prompt for vision-language models},
  author={Zhao, Cairong and Wang, Yubin and Jiang, Xinyang and Shen, Yifei and Song, Kaitao and Li, Dongsheng and Miao, Duoqian},
  journal={IEEE Transactions on Image Processing},
  volume={33},
  pages={1348--1360},
  year={2024},
  publisher={IEEE}
}

@InProceedings{yue2024mmmu,
  author    = {Yue, Xiang and Ni, Yuansheng and Zhang, Kai and Zheng, Tianyu and Liu, Ruoqi and Zhang, Ge and Stevens, Samuel and Jiang, Dongfu and Ren, Weiming and Sun, Yuxuan and others},
  booktitle = {Proceedings of the IEEE/CVF Conference on Computer Vision and Pattern Recognition},
  date      = {2024-06},
  title     = {Mmmu: A massive multi-discipline multimodal understanding and reasoning benchmark for expert agi},
  address   = {Seattle, WA, USA},
  doi       = {10.1109/cvpr52733.2024.00913},
  pages     = {9556--9567},
  publisher = {IEEE},
  year      = {2024},
}

@misc{lieneman2023,
  author = {Lieneman, Corey},
  title = {10 Therapist Coaches Big Ignore: Parent-Child Interaction Therapy (PCIT) for Older Children},
  year = {2023},
  month = {6},
  howpublished = {\url{https://www.youtube.com/watch?v=YUkujhg6j6w}},
  note = {Accessed: 2024-03-15}
}

@misc{adam2013,
  author = {Adam},
  title = {Piaget - Object permanence failure (Sensorimotor Stage)},
  year = {2013},
  month = {1},
  howpublished = {\url{https://www.youtube.com/watch?v=rVqJacvywAQ}},
  note = {Accessed: 2025-09-12}
}

@misc{Lieneman2024,
  author = {Lieneman, Corey},
  title = {5 Parent Meets CDI Goal Criteria: Parent-Child Interaction Therapy (PCIT) for Older Children},
  year = {2023},
  month = {6},
  howpublished = {\url{https://www.youtube.com/watch?v=N3wAPLXd7I0}},
  note = {Accessed: 2025-09-12}
}

@inproceedings{szymanski2025limitations,
  title={Limitations of the llm-as-a-judge approach for evaluating llm outputs in expert knowledge tasks},
  author={Szymanski, Annalisa and Ziems, Noah and Eicher-Miller, Heather A and Li, Toby Jia-Jun and Jiang, Meng and Metoyer, Ronald A},
  booktitle={Proceedings of the 30th International Conference on Intelligent User Interfaces},
  pages={952--966},
  year={2025}
}

@inproceedings{chen2024humans,
  title={Humans or LLMs as the Judge? A Study on Judgement Bias},
  author={Chen, Guiming and Chen, Shunian and Liu, Ziche and Jiang, Feng and Wang, Benyou},
  booktitle={Proceedings of the 2024 Conference on Empirical Methods in Natural Language Processing},
  pages={8301--8327},
  year={2024}
}

@inproceedings{dong2024can,
  title={Can LLM be a Personalized Judge?},
  author={Dong, Yijiang and Hu, Tiancheng and Collier, Nigel},
  booktitle={Findings of the Association for Computational Linguistics: EMNLP 2024},
  pages={10126--10141},
  year={2024}
}
